\documentclass[manuscript]{acmart}
\usepackage{xcolor}
\usepackage{xspace}
\usepackage{multicol}
\usepackage{multirow}
\usepackage{colortbl}
\usepackage[most]{tcolorbox}
\usepackage{subcaption}
\usepackage{graphicx}
\usepackage{tcolorbox}
\usepackage{framed}
\usepackage{changepage}
\usepackage{fontawesome5}
\usepackage{pgfplots}
\usepackage{pgfplotstable}

\AtBeginDocument{%
  }

\setcopyright{acmlicensed}
\copyrightyear{2018}
\acmYear{2018}
\acmDOI{XXXXXXX.XXXXXXX}
\acmConference[Conference acronym 'XX]{Make sure to enter the correct
  conference title from your rights confirmation email}{June 03--05,
  2018}{Woodstock, NY}
\acmISBN{978-1-4503-XXXX-X/18/06}

\begin{document}
\newcommand{\todo}[1]{{\color{orange} \bfseries TODO: #1}}
\newcommand{\jenny}[1]{{\color{purple} \bfseries Jenny: #1}}
\newcommand{\aayush}[1]{{\color{blue} \bfseries Aayush: #1}}
\newcommand{\ysrt}[1]{{\color{cyan} \bfseries Yasharth: #1}}
\newcommand{\added}[1]{{\color{black} #1}}
\newcommand{\aaydded}[1]{{\color{cyan} #1}}

\newcommand{\tool}{TableTalk\xspace}
\newcommand{\baseline}{baseline language agent\xspace}
\newcommand{\gpt}{\texttt{GPT-4o}\xspace}

\definecolor{PAblue}{RGB}{0,122,204}%
\definecolor{PAlightblue}{RGB}{235, 247, 255}%
\definecolor{color1}{RGB}{223, 246, 244}
\definecolor{green3}{RGB}{66,179,130}
\definecolor{green2}{RGB}{121,205,169}
\definecolor{green1}{RGB}{196,233,217}
\definecolor{red1}{RGB}{251,219,220}
\definecolor{red2}{RGB}{244,164,166}
\definecolor{red3}{RGB}{236,91,96}
\definecolor{gray1}{RGB}{220,220,220}
\definecolor{blue1}{RGB}{101,173,246}
\definecolor{blue2}{RGB}{12,112,212}
\definecolor{orange0}{RGB}{253, 238, 216}
\definecolor{orange1}{RGB}{250,181,97}
\definecolor{orange2}{RGB}{245,138,7}

\definecolor{boxcolor}{RGB}{238, 223, 204} %
\DeclareRobustCommand{\mybox}[2][gray!20]{%
\begin{tcolorbox}[   %
        breakable,
        left=0pt,
        right=0pt,
        top=0pt,
        bottom=0pt,
        colback=#1,
        colframe=black,
        width=\dimexpr\columnwidth\relax, 
        enlarge left by=0mm,
        boxsep=5pt,
        outer arc=4pt,
        boxrule=.5mm
        ]
        #2
\end{tcolorbox}
}

\newcommand{\theme}[2]{#2\xspace\emph{#1}}
\newtcbox{\formativecode}[1][]{enhanced,
 box align=base,
 nobeforeafter,
 colback=PAlightblue,
 colframe=PAlightblue,
 fontupper=\small\ttfamily,
 left=0.2pt,
 right=0.2pt,
 top=0.2pt,
 bottom=0.2pt,
 boxsep=0.4pt,
 #1}

\newtcbox{\evalcode}[1][]{enhanced,
 box align=base,
 nobeforeafter,
 colback=orange0,
 colframe=orange0,
 fontupper=\small\ttfamily,
 left=0.2pt,
 right=0.2pt,
 top=0.2pt,
 bottom=0.2pt,
 boxsep=0.4pt,
 #1}

 \newtcbox{\ilabel}[1][]{enhanced,
 box align=base,
 nobeforeafter,
 colback=PAblue,
 colframe=PAblue,
 size=small,
 fontupper=\color{white}\scriptsize\bf\sffamily,
 left=0.2pt,
 right=0.2pt,
 top=0.2pt,
 bottom=0.2pt,
 boxsep=2pt,
 arc=4.5pt,
 #1}

\newcommand{\pblockquote}[2]{\begin{quote}\emph{"{#1}"} (P#2)\end{quote}}
\newcommand{\pquote}[2]{\emph{"{#1}"} (P#2)}
\newcommand{\equote}[2]{\emph{"{#1}"} (E#2)}

\newcommand{\icon}[1]{{\includegraphics[height=1.5\fontcharht\font`\B]{#1}}\xspace}
\newcommand{\meiicon}{\icon{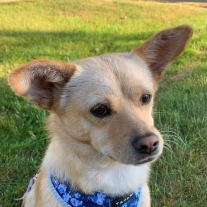}}

\newcommand{\dgA}{DP1: Scaffolding}

\newcommand{\dgB}{DP2: Flexibility}

\newcommand{\dgC}{DP3: Incrementality}

\newcommand{\fA}{F1: Following a plan based on the process of experts}
\newcommand{\fB}{F2: Suggesting three next steps in the plan to the programmer for human-in-the-loop planning}
\newcommand{\fC}{F3: Rapid spreadsheet table prototyping by previewing tables in Markdown}
\newcommand{\fD}{F4: Tools to build atomic components of spreadsheets}

\newcommand{\scenario}[1]{ 
	\vspace{-0.15cm}
	\def\FrameCommand{%
		\hspace{0pt}%
		{\color{PAblue}\vrule width 2pt}%
		{\color{white}\vrule width 2pt}%
		\colorbox{white}
	}%
	\MakeFramed{\advance\hsize-\width\FrameRestore}%
	\noindent\hspace{-4.55pt}%
	\begin{adjustwidth}{}{0pt}
		\emph{#1}
		\vspace{-3pt}
	\end{adjustwidth}\endMakeFramed%
}

\def\polaritybarchart#1#2#3#4#5{
\resizebox{0.08\linewidth}{7.5pt} {
\begin{tikzpicture}[]
\node[] { \huge \emph{#4}};
\end{tikzpicture}
}
\resizebox {0.81\linewidth} {6.5pt} {%
\begin{tikzpicture}[]
\begin{axis}[
      axis background/.style={fill=gray!30, draw=gray!30},
      axis line style={draw=none},
      tick style={draw=none},
      ytick=\empty,
      xtick=\empty,
      ymin=0, ymax=0.70,
      xmin=0, xmax=3]
\addplot [
      ybar interval=.5,
      fill=green3,
      draw=none,
]
	coordinates {(3*#1,1) (0,0.30)}; %
\addplot [
      ybar interval=.5,
      fill=red2,
      draw=none,
]
	coordinates {(3*(#1+#2),1) (3*#1,1)}; %
\addplot [
      ybar interval=.5,
      fill=gray1,
      draw=none,
]
	coordinates {(3*(#3+#2+#1),1) (3*(#2+#1),1)}; %
\end{axis}%
\end{tikzpicture}%
}
\resizebox{0.08\linewidth}{7.5pt} {
\begin{tikzpicture}[]
\node[] { \huge \emph{#5}};
\end{tikzpicture}
}
}

\def\mylegend#1#2{
\resizebox {0.02\linewidth} {6.5pt} {%
\begin{tikzpicture}[]
\begin{axis}[
      axis background/.style={fill=white!30, draw=white!30},
      axis line style={draw=none},
      tick style={draw=none},
      ytick=\empty,
      xtick=\empty,
      ymin=0, ymax=0.70,
      xmin=0, xmax=6]
\addplot [
      ybar interval=.5,
      fill=#2,
      draw=none,
]
	coordinates {(4.5,1) (0,0.30)}; %
\end{axis}%
\end{tikzpicture}%
}%
#1
}

\newcommand{\mylabel}[1]{{\small{\emph{#1}}}}

\def\cqbarchart#1#2#3#4#5#6#7#8{
\resizebox{0.08\linewidth}{7.5pt} {
\begin{tikzpicture}[]
\node[] { \huge \emph{#7}};
\end{tikzpicture}
}
\resizebox {0.81\linewidth} {6.5pt} {%
\begin{tikzpicture}[]
\begin{axis}[
      axis background/.style={fill=gray!30, draw=gray!30},
      axis line style={draw=none},
      tick style={draw=none},
      ytick=\empty,
      xtick=\empty,
      ymin=0, ymax=0.70,
      xmin=0, xmax=6]
\addplot [
      ybar interval=.5,
      fill=green3,
      draw=none,
]
	coordinates {(6*#1,1) (0,0.30)}; %
\addplot [
      ybar interval=.5,
      fill=green2,
      draw=none,
]
	coordinates {(6*(#1+#2),1) (6*#1,1)}; %
\addplot [
      ybar interval=.5,
      fill=gray1,
      draw=none,
]
	coordinates {(6*(#3+#2+#1),1) (6*(#2+#1),1)}; %
\addplot [
      ybar interval=.5,
      fill=red2,
      draw=none,
]
	coordinates {(6*(#4+#3+#2+#1),1) (6*(#3+#2+#1),1)}; %
\addplot [
      ybar interval=.5,
      fill=red3,
      draw=none,
]
	coordinates {(6*(#5+#4+#3+#2+#1),1) (6*(#4+#3+#2+#1),1)}; %
\addplot [
      ybar interval=.5,
      fill=red1,
      draw=none,
]
	coordinates {(6*(#6+#5+#4+#3+#2+#1),1) (6*(#5+#4+#3+#2+#1),1)}; %
\end{axis}%
\end{tikzpicture}%
}
\resizebox{0.08\linewidth}{7.5pt} {
\begin{tikzpicture}[]
\node[] { \huge \emph{#8}};
\end{tikzpicture}
}
}

\def\tlxbarchart#1#2#3#4#5#6#7#8{
\resizebox{0.08\linewidth}{7.5pt} {
\begin{tikzpicture}[]
\node[] { \huge \emph{#7}};
\end{tikzpicture}
}
\resizebox {0.81\linewidth} {6.5pt} {%
\begin{tikzpicture}[]
\begin{axis}[
      axis background/.style={fill=gray!30, draw=gray!30},
      axis line style={draw=none},
      tick style={draw=none},
      ytick=\empty,
      xtick=\empty,
      ymin=0, ymax=0.70,
      xmin=0, xmax=6]
\addplot [
      ybar interval=.5,
      fill=blue2,
      draw=none,
]
	coordinates {(6*#1,1) (0,0.30)}; %
\addplot [
      ybar interval=.5,
      fill=blue1,
      draw=none,
]
	coordinates {(6*(#1+#2),1) (6*#1,1)}; %
\addplot [
      ybar interval=.5,
      fill=gray1,
      draw=none,
]
	coordinates {(6*(#3+#2+#1),1) (6*(#2+#1),1)}; %

\addplot [
      ybar interval=.5,
      fill=orange1,
      draw=none,
]
	coordinates {(6*(#4+#3+#2+#1),1) (6*(#3+#2+#1),1)}; %
\addplot [
      ybar interval=.5,
      fill=orange2,
      draw=none,
]
	coordinates {(6*(#5+#4+#3+#2+#1),1) (6*(#4+#3+#2+#1),1)}; %
\addplot [
      ybar interval=.5,
      fill=orange2,
      draw=none,
]
	coordinates {(6*(#6+#5+#4+#3+#2+#1),1) (6*(#5+#4+#3+#2+#1),1)}; %
\end{axis}%
\end{tikzpicture}%
}
\resizebox{0.08\linewidth}{7.5pt} {
\begin{tikzpicture}[]
\node[] { \huge \emph{#8}};
\end{tikzpicture}
}
}

\title{\tool: Scaffolding Spreadsheet Development with a Language Agent}

\newcommand{\equalSupervision}{\authornote{Denotes equal supervision. Author names are listed alphabetically. \newline \textsuperscript{\dag}Work completed while at Microsoft.}}

\author{Jenny T. Liang}
\authornotemark[2]
\email{jtliang@cs.cmu.edu}
\affiliation{%
  \institution{Carnegie Mellon University}
  \city{Pittsburgh}
  \state{PA}
  \country{USA}
}

\author{Aayush Kumar}
\email{t-aaykumar@microsoft.com}
\affiliation{%
  \institution{Microsoft}
  \city{Banglore}
  \country{India}
}

\author{Yasharth Bajpai}
\email{ybajpai@microsoft.com}
\affiliation{%
  \institution{Microsoft}
  \city{Banglore}
  \country{India}
}

\author{Sumit Gulwani}
\email{sumitg@microsoft.com}
\affiliation{%
  \institution{Microsoft}
  \city{Redmond}
  \city{WA}
  \country{USA}
}

\author{Vu Le}
\email{levu@microsoft.com}
\affiliation{%
  \institution{Microsoft}
  \city{Redmond}
  \city{WA}
  \country{USA}
}

\author{Chris Parnin}
\affiliation{%
  \institution{Microsoft}
  \city{Redmond}
  \city{WA}
  \country{USA}
}
\email{chrisparnin@microsoft.com}

\author{Arjun Radhakrishna}
\affiliation{%
  \institution{Microsoft}
  \city{Redmond}
  \city{WA}
  \country{USA}
}
\email{arradha@microsoft.com}

\author{Ashish Tiwari}
\affiliation{%
  \institution{Microsoft}
  \city{Redmond}
  \city{WA}
  \country{USA}
}
\email{ashish.tiwari@microsoft.com}

\author{Emerson Murphy-Hill}
\equalSupervision
\affiliation{%
  \institution{Microsoft}
  \city{Redmond}
  \city{WA}
  \country{USA}
}
\email{emerson.rex@microsoft.com}

\author{Gustavo Soares}
\authornotemark[1]
\email{gustavo.soares@microsoft.com}
\affiliation{%
  \institution{Microsoft}
  \city{Redmond}
  \city{WA}
  \country{USA}
}

\renewcommand{\shortauthors}{Liang et al.}

\begin{abstract}
Spreadsheet programming is challenging.
Programmers use spreadsheet programming knowledge (e.g., formulas) and problem-solving skills to combine actions into complex tasks. 
Advancements in large language models have introduced language agents that observe, plan, and perform tasks, showing promise for spreadsheet creation. 
We present TableTalk, a spreadsheet programming agent embodying three design principles—scaffolding, flexibility, and incrementality—derived from studies with seven spreadsheet programmers and 85 Excel templates.
TableTalk guides programmers through structured plans based on professional workflows, generating three potential next steps to adapt plans to programmer needs. 
It uses pre-defined tools to generate spreadsheet components and incrementally build spreadsheets. 
In a study with 20 programmers, TableTalk produced higher-quality spreadsheets 2.3 times more likely to be preferred than the baseline.
It reduced cognitive load and thinking time by 12.6\%. 
From this, we derive design guidelines for agentic spreadsheet programming tools and discuss implications on spreadsheet programming, end-user programming, AI-assisted programming, and human-agent collaboration.
\end{abstract}


\received{20 February 2007}
\received[revised]{12 March 2009}
\received[accepted]{5 June 2009}

\maketitle

\begin{figure}[t!]
\centering
\includegraphics[trim=0 350 200 0, clip, width=\linewidth, keepaspectratio]{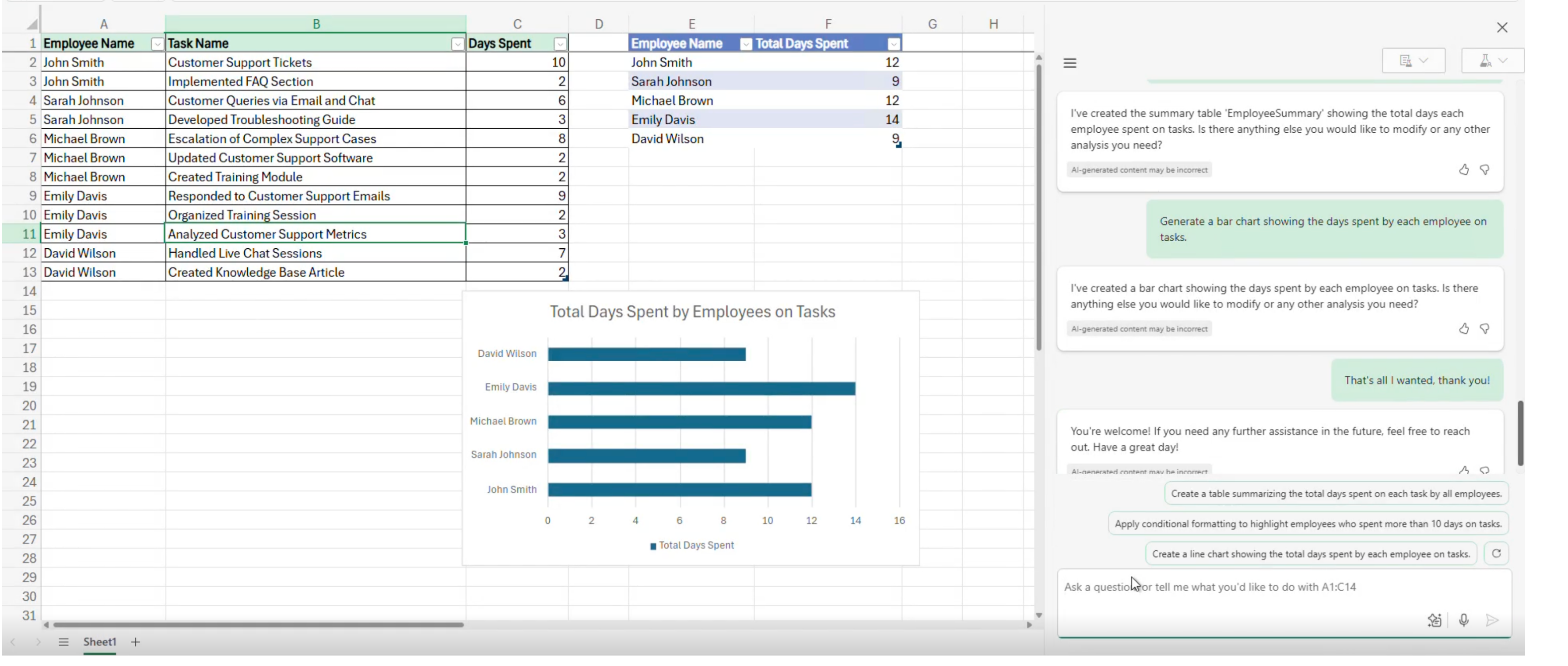} 
\caption{
A screenshot of the \tool interface.
}
\label{fig:intro}
\end{figure}

\section{Introduction}
\label{sec:introduction}

\scenario{
Alma is creating a spreadsheet to track expenses to understand last month's spending trends of her business with her co-owner.
Alma is unsure on how to solve this problem, so she begins by defining a table whose schema represents the information she wants to track. 
This includes columns for the transaction date, cost, and spending category.
Then, she manually tabulates the data from various sources, such as bank statements or receipts.
Alma now wants to understand her top spending categories from the past month, which requires completing multiple steps.
First, Alma must produce a second table by filtering rows from the past month, aggregating the rows by spending category, summing their costs, and saving the results in a second table.
She then must sort the second table by the total cost column.
However, Alma does not know how to decompose the problem into these steps.
She also struggles to implement the formula due to having difficulty understanding complex spreadsheet programming APIs, and eventually gives up on implementing the spreadsheet.
}

Like Alma, many professionals develop spreadsheet programs~\cite{scaffidi2005estimating, abraham2008spreadsheet, ko2011state}, from administrators and retail managers~\cite{glass2024skills} to teachers and analysts~\cite{ko2011state}.
We refer to these professionals as \emph{spreadsheet programmers}.
Spreadsheet programming is an important skill in the modern workforce~\cite{scaffidi2005estimating, scaffidi2017workers, glass2024skills}.
Based on a 2023 longitudinal analysis of millions of job postings, spreadsheet programming remains a significant and enduring skill~\cite{glass2024skills}.
Despite the prevalence of spreadsheet programs, creating these programs is challenging.
As illustrated by Alma's experience, what makes spreadsheet programming challenging is two-fold.
Firstly, spreadsheet programming benefits from having some programming expertise~\cite{nardi1991twinkling} and prerequisite understanding of APIs~\cite{ko2011state} to perform actions, like writing formulas.
Secondly, spreadsheet programming can require a degree of problem-solving knowledge to understand how to approach the problem and decompose the task into individual actions that achieve the intended goal. 
Programmers often achieve this by developing and following systematic processes~\cite{pirolli2005sensemaking, latoza2020explicit, arab2022exploratory}.
Without adequate support for these kinds of knowledge, spreadsheet programmers struggle to understand and manage spreadsheets~\cite{reschenhofer2015empirical} and introduce errors into the program~\cite{abraham2008spreadsheet}.

The emergence of large language models (LLMs), such as GPT-4~\cite{achiam2023gpt}, offer a promising avenue to assist spreadsheet programmers since they address both of the aforementioned challenges.
These models have sufficient spreadsheet programming knowledge to complete individual spreadsheet programming tasks (e.g., formula prediction~\cite{joshi2024flame, chen2021spreadsheetcoder}).
Further, LLMs can solve complex tasks via strong planning capabilities developed through advanced reasoning techniques~\cite{shinn2024reflexion, yao2024tree}, which have enabled LLM-based \emph{language agents} to become effective planners and problem solvers for spreadsheet programming~\cite{li2024sheetcopilot}. 
Compared to traditional LLMs, which focus on generating text, language agents can also collect observations of the environment and plan and perform actions to update their environment~\cite{andreas2022language, wang2024survey, xi2023rise}.

Humans and language agents assume complementary roles in completing tasks.
Language agents can scaffold and guide humans through complex tasks by following expert processes~\cite{sun2024reviewflow, bajpai2024lets}.
At the same time, research shows that language agents benefit from human intelligence, such as learning from human problem-solving processes~\cite{wang2024survey, bajpai2024lets}.
Therefore, humans could benefit language agents by improving the agent's planning and task performance via human expertise, while language agents could guide humans through expert processes to complete complex tasks the individual would struggle with on their own.
This motivates the need to study human-agent interaction from a planning and problem-solving perspective, which remains understudied.
Better understanding how humans and language agents interact could improve their alignment and could also inform our understanding of usable agent systems, which is vital for useful and practical tools~\cite{myers2016programmers}.This becomes increasingly important given the growing capabilities of agents to complete a wider range of tasks (e.g., software engineering~\cite{li2024sheetcopilot, yang2024swe} or real-life scenarios~\cite{xi2023rise}) and the adoption of agents in commercial programming systems like Devin~\cite{devin2025devin}.

In this work, we explore the potential for collaboration between language agents and humans to plan and solve complex tasks together.
We use spreadsheet programming as a domain to study human-agent collaboration, given the ability of language agents to perform the task~\cite{li2024sheetcopilot}; the semi-structured nature of the task~\cite{pirolli2005sensemaking}, indicating the need for planning and problem solving by spreadsheet programmers; and the difficulty of the task for humans~\cite{ko2011state}.
We explore this idea with our tool, \tool (see Figure~\ref{fig:intro}).
\tool implements three design principles for tools to support spreadsheet programming---\emph{scaffolding}, \emph{flexibility}, and \emph{incrementality}.
We derive these principles by understanding the design space of scaffolding tools and how programmers use them through a study of 85 spreadsheet templates and a user study with 7 spreadsheet programmers respectively.
\tool supports the creation of spreadsheet programs by having the programmer participate in the planning process and provide feedback on potential table schema through a chat interface.
\tool is implemented as a language agent that 1) follows a plan for effective spreadsheet creation identified by \citet{pirolli2005sensemaking} and 2) has access to tools to implement atomic spreadsheet features, enabling the incremental development of spreadsheets.
In each interaction with \tool, the tool generates three potential next steps that the programmer can select to adapt the plan to their context.
We demonstrate the benefits of the design principles via a user study with 20 participants.
We find that \tool produces higher-quality spreadsheets that are 2.3 times more likely to be preferred by other spreadsheet programmers compared to a \baseline's.
It also enables better conversation, reduces cognitive load, and decreases the amount of time spent thinking about spreadsheet programming actions by 1.9 minutes (112 seconds)---12.6\% of the total task time---compared to the baseline.

In summary, the contributions of this work are:
\begin{itemize}
    \item A list of design principles (Section~\ref{sec:design-goals}) for tools to support spreadsheet programming based on a formative study of 85 spreadsheet templates (Section~\ref{sec:spreadsheet-template-analysis}) and a user study with 7 spreadsheet programmers (Section~\ref{sec:formative-user-study});
    \item \tool, a novel system that reifies these principles to provide iterative, flexible, and scaffolded guidance for programmers to build spreadsheets in Excel with a language agent (Section~\ref{sec:tool});
    \item Insights on AI-assisted spreadsheet programming and human-agent collaboration via an evaluation of \tool with 20 spreadsheet programmers (Section~\ref{sec:evaluation-user-study}); and
    \item Design guidelines and recommendations for future human-agent collaborative systems (Section~\ref{sec:discussion}).
\end{itemize}

\section{Related Work}
Below, we discuss prior work on human factors of spreadsheet programming (Section~\ref{sec:human-factors-of-spreadsheet-programming}), scaffolding tools for programming tasks (Section~\ref{sec:scaffolding-processes}), and AI for spreasheet programming (Section~\ref{sec:llm-agents}).
Given the rapid advances in LLMs, our discussions offer a snapshot of the field as of December 2024.

\subsection{Human Factors of Spreadsheet Programming}
\label{sec:human-factors-of-spreadsheet-programming}
Prior work has investigated the process that spreadsheet programmers follow to create spreadsheets, revealing its complex and iterative nature.
In a field study of 11 spreadsheet programmers, \citet{nardi1991twinkling} found that spreadsheet development was collaborative by nature and required both programming and domain expertise.
This expertise was often distributed among multiple individuals who had different specialties.
Closely related is \citet{pirolli2005sensemaking}'s cognitive task analysis, which found that professionals who developed knowledge products from data like spreadsheets followed a defined process.
This involved gathering information, developing a schema for analysis, generating insights by manipulating the schema, and creating some knowledge product or action.
Further, the authors found that the expert process involved a foraging loop (i.e., finding, searching, filtering, and extracting information) and a sense making loop (i.e., developing a mental model for a schema).

Other studies have revealed the challenges associated with spreadsheet development.
Spreadsheet programmers often struggle to find the correct abstractions to use and to understand how to reuse existing code to accomplish a task~\cite{ko2011state}.
In addition, spreadsheet programs can be difficult to understand, extend, and manage~\cite{reschenhofer2015empirical}. 
As a result, they can be riddled with errors, including mechanical and logic errors~\cite{abraham2008spreadsheet}.
Although automating aspects of spreadsheet programming could be helpful~\cite{reschenhofer2015empirical}, it can also come at a cost to the programmer.
\citet{pandita2018no} found that there was no clear benefit in using tools in spreadsheet programming: while tool usage could increase task correctness, it could also result in longer task completion times.

The prior literature on human factors of spreadsheet programming reveals rich insights on how programmers develop spreadsheets.
However, these works do not study how spreadsheet programmers use scaffolding tools like spreadsheet templates, which is vital to understanding how AI can guide spreadsheet development.
We corroborate and extend the findings of these studies by examining the benefits and challenges of using scaffolding tools in spreadsheet development in our formative user study (see Section~\ref{sec:formative-user-study}).
In addition, we leverage the insights from these studies in the design principles (see Section~\ref{sec:design-goals}) and implementation of \tool (see Section~\ref{sec:tool}),
such as having the agent follow the process from \citet{pirolli2005sensemaking} for spreadsheet programmers.

\subsection{Scaffolding Tools for Programming}
\label{sec:scaffolding-processes}
Scaffolding tools help users complete complex tasks by structuring tasks in a way such that they can accomplish tasks that they struggle to complete on their own~\cite{reiser2018scaffolding}.
Examples of scaffolding tools include templates, examples of other work, hints, or links to resources~\cite{saye2002scaffolding}, as well as step-by-step instructions to complete a task~\cite{latoza2020explicit}.
In programming, scaffolding tools can help programmers with problem solving and planning. 
They have been applied to a variety of complex programming tasks, such as decision making~\cite{liu2019unakite}, debugging~\cite{bajpai2024lets, latoza2020explicit}, software architecture, and testing~\cite{arab2021howtoo}.
However, scaffolding tools struggle to offer the right level of support by being too general or too specific~\cite{reiser2018scaffolding}.
For example, when a programmer follows expert-defined instructions to solve programming problems---known as programming strategies~\cite{latoza2020explicit}---it can be difficult to apply the strategy because it may not consider the programmer's unique context or be written too abstractly~\cite{arab2022exploratory}.

LLMs have shown promise in alleviating these challenges by imbuing additional flexibility and user context into scaffolding tools for complex tasks, such as for the academic peer review process~\cite{sun2024reviewflow} and information sense making~\cite{suh2023sensecape, liu2024selenite}.
In programming, LLM-based scaffolding approaches have shown promise in helping programmers.
The most closely related tool to \tool is ROBIN~\cite{bajpai2024lets}, a tool that guides programmers through a predefined debugging programming strategy.
ROBIN is implemented with four main LLM-based components, including one for generating follow-ups in the conversation.
With this approach, ROBIN led to a 150\% increase in defect localization rates and a 250\% increase in defect resolution rates compared to the baseline.

Overall, this body of literature points to the promise of scaffolding tools assisting spreadsheet programmers in creating spreadsheets.
However, to the best of our knowledge, scaffolding tools for spreadsheet development remain understudied.
This is because, rather than guiding programmers through complex tasks, many intelligent spreadsheet tools have focused on developing improved interfaces, algorithms, or machine learning techniques for a range of spreadsheet development tasks.
This includes writing formulas~\cite{srinivasa2022gridbook}, prototyping tables~\cite{huang2024table}, debugging~\cite{abraham2007goaldebug, myers1991graphical} as well as defining test cases~\cite{burnett2002testing}, assertions~\cite{burnett2003end}, and constraints~\cite{myers1991graphical}.
Of these tools, perhaps the most related to \tool is Table Illustrator~\cite{huang2024table} and GridBook~\cite{srinivasa2022gridbook}.
Table Illustrator~\cite{huang2024table} is an interactive system that facilitates the rapid prototyping of different table layouts by allowing programmers to arrange puzzle pieces as a metaphor for table construction.
In the Table Illustrator user study, \citet{huang2024table} found that the tool performed similarly to Excel, while decreasing cognitive load and task completion time.
Meanwhile, GridBook~\cite{srinivasa2022gridbook} is an interactive system that allows programmers to provide input in natural language within a spreadsheet cell and uses deep learning techniques to generate formulas.
In the GridBook user study, \citet{srinivasa2022gridbook} found that the tool performed comparably to Excel, but reduced task completion time compared to Jupyter notebooks.

However, some work has studied scaffolding spreadsheet programming, namely, detecting and producing spreadsheet templates~\cite{abraham2004header, abraham2006inferring, abraham2005visual, erwig2006gencel}.
ViTSL~\cite{abraham2005visual} is a visual programming language that defines abstractions for the spreadsheet structure to reduce spreadsheet programming errors.
It offers abstractions on cells, references, vertical groups, and horizontal groups.
Gencel~\cite{erwig2006gencel}, an Excel add-on, then generates and edits tables that conform to the ViSTL template structure.
In Gencel's user study, \citet{erwig2006gencel} found that the participants wanted more flexibility in ViSTL templates and expressed concern in being able to adapt ViSTL templates due to the required programming knowledge.

To our knowledge, \tool is the first conversational tool that uses LLM language agents to scaffold the spreadsheet creation process.
Compared to Table Illustrator~\cite{huang2024table} and GridBook~\cite{srinivasa2022gridbook}, \tool follows a scaffolding approach to develop spreadsheets. 
Unlike GridBook, \tool uses conversation rather than annotations in spreadsheet cells to develop spreadsheets and further, can manipulate spreadsheet presentation (e.g., themes) in addition to complex formulas; unlike Table Illustrator, \tool uses state-of-the-art LLMs to create spreadsheets and supports the development of complex formulas, rather than being limited to simple summations.
Next, we build on previous work on spreadsheet programming scaffolding tools, such as ViSTL~\cite{abraham2005visual} and Gencel~\cite{erwig2006gencel}, by using language agents to provide additional flexibility in the spreadsheet creation process.
Finally, we extend the existing body of literature on LLM scaffolding tools such as ROBIN~\cite{bajpai2024lets} by studying the collaboration between programmers and agents that can modify the development environment, which is becoming increasingly important with the adoption of agentic approaches in practice with 
commercial tools like Devin~\cite{devin2025devin}. 

\subsection{AI for Spreadsheet Programming}
\label{sec:llm-agents}
Prior work has investigated multiple approaches for LLMs to solve spreadsheet programming tasks.
Previously, LLMs solved a range of self-contained spreadsheet programming tasks through pre-training and fine-tuning, such as formula prediction~\cite{joshi2024flame, chen2021spreadsheetcoder} and cell role prediction~\cite{du2021tabularnet}.
Recently, a budding approach has allowed LLMs to complete more complex tasks and increase task performance in spreadsheet programming: \emph{language agents}, also known as agents.
Agents achieve a specific goal by observing the environment, performing actions, generating utterances, modeling internal state, and inferring intentions from others~\cite{andreas2022language, xi2023rise, wang2024survey}.
Compared to traditional LLMs, which focus on generating text autoregressively (i.e., from left to right) given the previous text, agents put special emphasis on dynamic planning and proactive action-taking within an environment to complete tasks.
In software engineering, language agents have grown rapidly in popularity \cite{casper2025aiagentindex}, achieving state-of-the-art performance in a variety of software development tasks and benchmarks \cite{jimenez2024swebenchlanguagemodelsresolve, zhang2025swebenchgoeslive}.
Such agents have appeared both in research \cite{yang2024swe, tufano2024autodev, autocoderover, wang2025openhandsopenplatformai}, as well as in commercial tools such as Cursor AI~\cite{cursor2025cursor}, Devin~\cite{devin2025devin}, Visual Studio Code Agent Mode~\cite{vscode_agent_mode} and GPT Pilot~\cite{github2025gptpilot}.
Notably, coding language agents differ from traditional AI programming assistants that only provide code autocompletions or suggestions. 
The former can provide proactive coding assistance in the development environment, while the latter lack the ability to make direct edits to code or iterate on their outputs.
Following the trend of coding language agents, \citet{li2024sheetcopilot} developed an agent to manipulate spreadsheets called SheetCopilot. 
SheetCopilot can understand spreadsheet manipulation requests in natural language and construct and execute plans to achieve the request.
Based on a benchmark of 221 spreadsheet programming tasks, SheetCopilot completed 44.3\% of the tasks in a single generation.

Although these advances have enabled state-of-the-art performance, agentic systems do not typically consider the role of human interaction, as they restrict human involvement to providing annotations on agent actions~\cite{xi2023rise} or limiting interactions to the beginning of the task rather than throughout the task (e.g.,~\cite{yang2024swe, li2024sheetcopilot}).
This reduces the usability of agent-based systems, as these approaches violate long-established principles in human-AI collaboration that underscored the importance of granular user feedback~\cite{amershi2019guidelines} and the refinement of AI output~\cite{horvitz1999principles}.
Thus, compared to other agentic approaches, \tool explores human-agent collaboration by being designed to work interactively with spreadsheet programmers to perform their tasks.
Rather than trying to complete a task in a single user interaction as in previous work~\cite{li2024sheetcopilot, yang2024swe, tufano2024autodev}, \tool scaffolds the spreadsheet creation process and allows programmers---rather than models---to provide feedback on intermediate states and evaluate multiple next steps for planning.

\section{Spreadsheet Template Study}
\label{sec:spreadsheet-template-analysis}
To better understand existing scaffolding tools in spreadsheet programming, we analyzed 85 Excel spreadsheet templates.
We studied spreadsheet templates because they are tools to scaffold the spreadsheet creation process for programmers.
Therefore, to understand spreadsheet templates analytically, our research question in this study is:

\begin{description}
\setlength{\itemsep}{0pt}
\setlength{\parskip}{2pt}
    \item[\textbf{RQ1}] What is the design space of Excel spreadsheet templates?
\end{description}

\label{sec:template-analysis}
We complement \citet{huang2024table}'s analysis on the design space of LaTeX and Microsoft Word tables by studying Excel spreadsheet templates.
Through a broad sample of 2,500 tables across a variety of domains, their analysis revealed rich insights into the different physical aspects of individual tables, including the structure and style of tables.
Our study extends this work by understanding the diverse problem-solving approaches that programmers employ to build spreadsheets.
Therefore, to answer RQ1, we gathered a dataset of Excel spreadsheet templates (Section~\ref{sec:template-analysis-data}) and qualitatively analyzed it (Section~\ref{sec:template-analysis-analysis}). 
We used the results (Section~\ref{sec:template-analysis-results}) of this study to inform the design principles of \tool and understand what spreadsheet components to generate.
All materials associated with the spreadsheet template study, including the datasets and codebooks, are included in the supplemental materials~\cite{supplemental-materials}.

\subsection{Data}
\label{sec:template-analysis-data}
For this study, we analyzed Excel templates since they are widely accessible scaffolding tools for spreadsheet programmers and are a sample of spreadsheets of higher quality and intended for public use.
These spreadsheets were representative of what we wanted \tool to produce.
For this analysis, we began by studying a single domain of spreadsheet templates to develop the codebook.
We then applied the codebook to more spreadsheet templates from a range of domains for generalizability.
The budget domain was selected as the initial domain to develop the codebook as it was a prominent predefined category of spreadsheet templates in the Excel application.

To obtain the Excel template data set, we used the Excel template library from Excel version 2409 and selected  the templates under the "Budget" filter.
This yielded 91 templates.
Because this set included spreadsheets that were not purely budgets (e.g., planners), we further filtered this set down by only looking at templates that contained the keyword "budget".
This produced the an initial budget dataset of 62 spreadsheets, which were then manually extracted for analysis.
To improve generalizability of the resulting taxonomy, we also collected a general template dataset containing the default Excel templates offered in Excel version 2506.
These templates spanned a variety of domains, such as task tracking, daily schedules, workout logs, and video game trackers.
After de-duplicating the data for any templates in the budget template dataset, the general template dataset contained 23 spreadsheets.

\subsection{Analysis}
\label{sec:template-analysis-analysis}
To analyze the Excel template dataset, two authors performed open coding on the dataset.
The authors independently coded 15 spreadsheets (approximately 25\% of the samples from the budget template dataset) for common atomic spreadsheet features and inductively generated codes in an individual codebook.
The authors then gathered to merge their individual codebooks.
Codes with similar themes were first merged.
Each of the remaining codes were added or removed based on unanimous agreement. 
The authors then grouped the codes into similar themes in a round of axial coding.
Each category was created only by unanimous agreement.

To validate the codebook, one author coded an additional spreadsheet by selecting areas within the spreadsheet and applying a code from the shared codebook. 
The codes were then masked, with the original highlighted areas corresponding to the code remaining intact.
The other author then re-coded the same spreadsheets by applying the shared codebook.
We then computed an inter-rater reliability (IRR) score.
Following recommendations in qualitative research~\cite{halpin2024inter}, we used Cohen's $\kappa$ for IRR as the codes distribution was not very skewed and there were two coders.
We did not use other metrics like Krippendorf's $\alpha$ and Gwet's AC1 and AC2, which is recommended for more larger and complex datasets with three or more coders~\cite{halpin2024inter}.
We obtained a Cohen's $\kappa$ score of 0.87---near perfect agreement~\cite{landis1977measurement}.
After the codebook was validated, the first author coded the remaining spreadsheets in the budget template dataset and repeated the process for the 23 spreadsheets in the general template dataset.
No new codes were discovered in this process.

\begin{figure}[t!]
\centering
\includegraphics[trim=0 250 725 0, clip, width=0.85\linewidth, keepaspectratio]{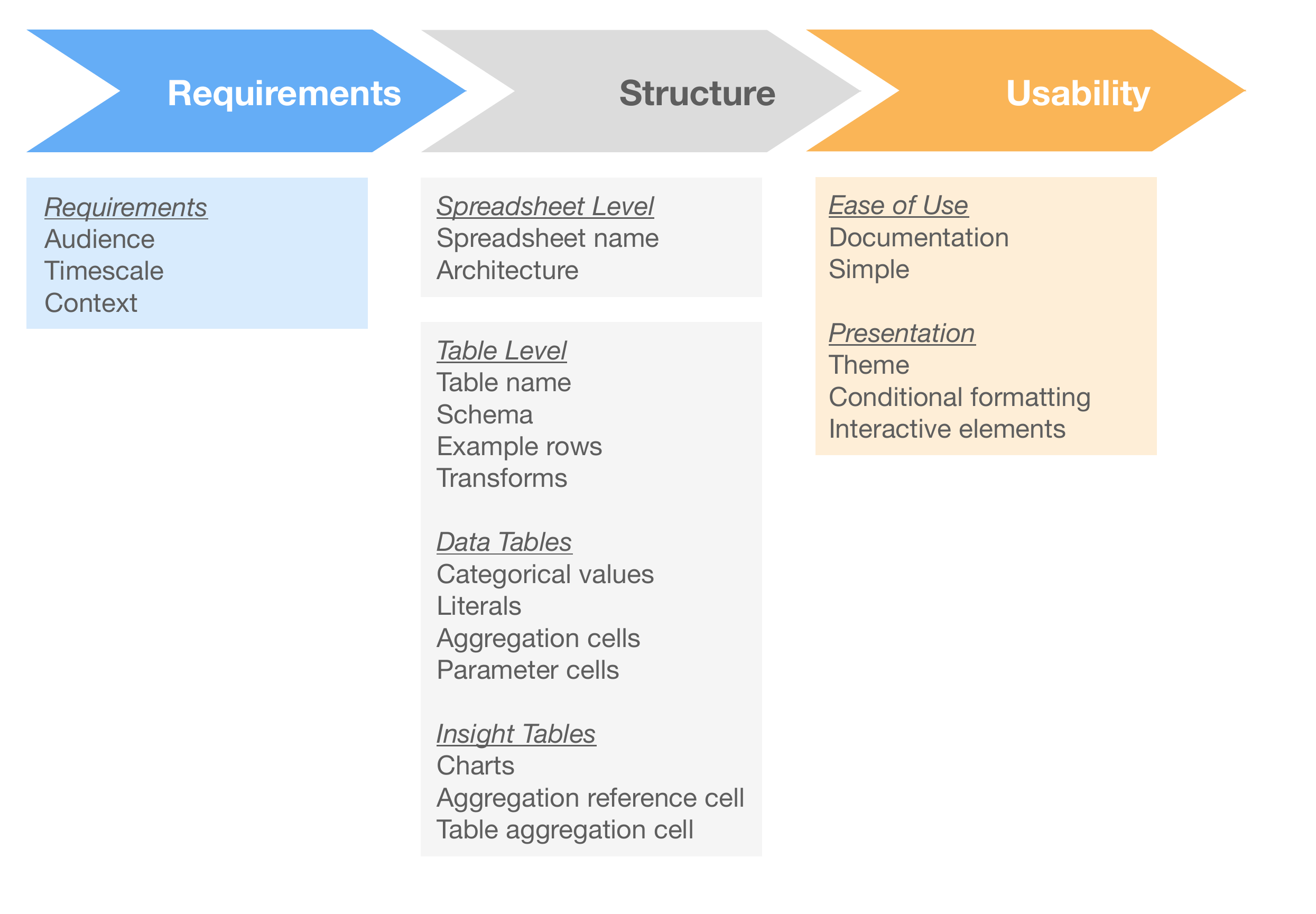} 
\caption{
An overview of the themes and codes of Excel spreadsheet template features from the spreadsheet template study.
The requirements of the spreadsheet (left) influence its physical structure (middle), defining the architecture and data organization.
This structure, in turn, serves as the foundation for its usability (right).
The themes associated with the requirements, structure, and usability are \underline{underlined}, with their corresponding codes listed directly below.
}
\label{fig:template-results-figure}
\end{figure}

\begin{figure}[t!]
\centering
\includegraphics[trim=0 25 700 0, clip, width=\linewidth, keepaspectratio]{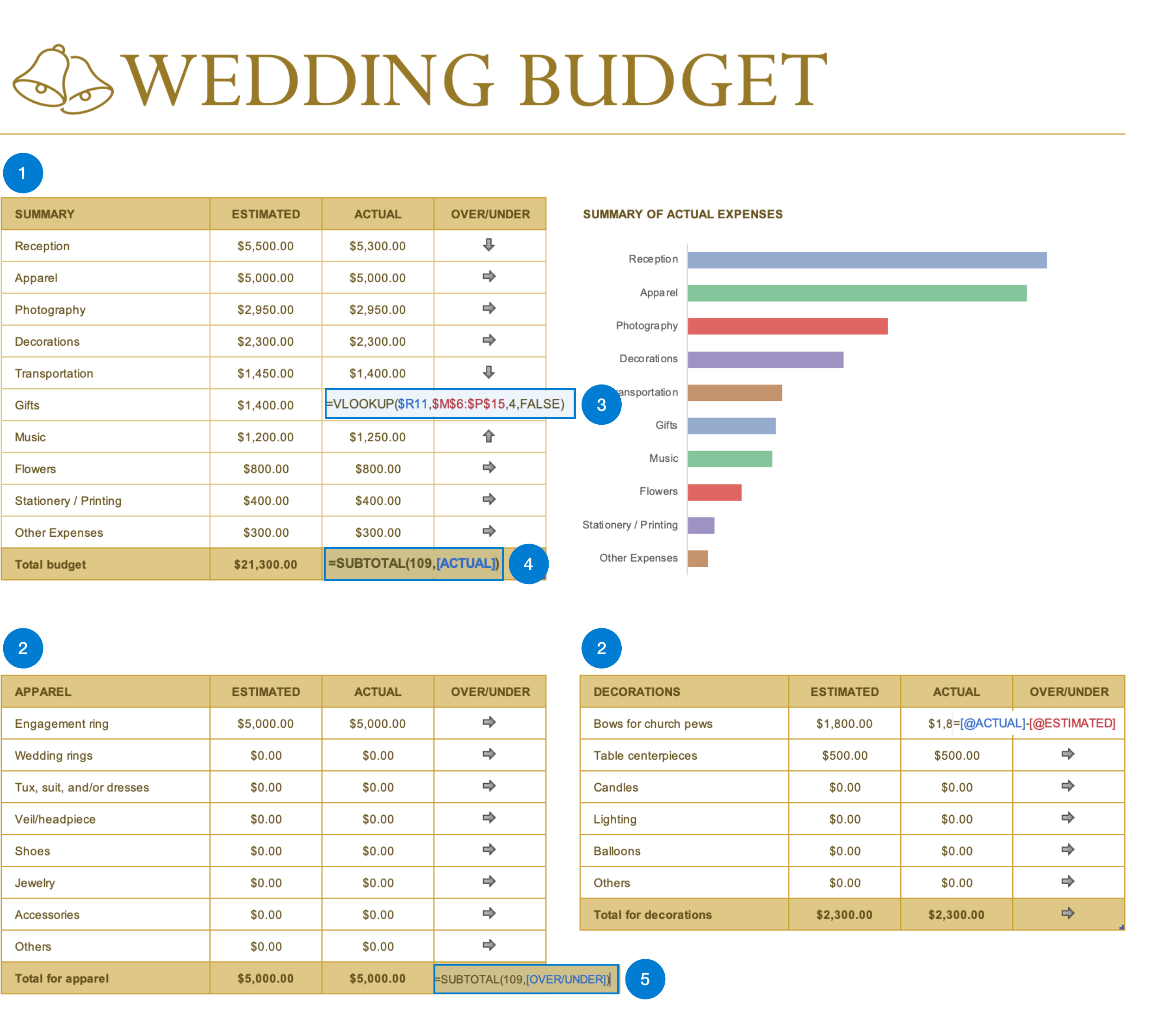} 
\caption{
An example Excel budget spreadsheet template, rearranged due to space constraints.
This template contains one \ilabel{1} \theme{insight table}{\faLightbulb[regular]}, which summarizes the data in the sheet in a table and visualization, and two \ilabel{2} \theme{data tables}{\faDatabase} that define the underlying data within the sheet. 
Within the data table, the last row is an \ilabel{5} \formativecode{aggregation cell}, which compiles the data within the single table.
Within the insight table, each row contains an \ilabel{3} \formativecode{aggregation reference cell} linking to the category's overall total via the aggregation cell from the corresponding data table.
The aggregation reference cells are then aggregated in the \ilabel{4} \formativecode{table aggregation cell} to summarize the costs across all the tables in the sheet.
}
\label{fig:template}
\end{figure}

\subsection{Results}
\label{sec:template-analysis-results}
Based on the qualitative analysis, we extracted 21 codes on common atomic features from Excel spreadsheet templates that were grouped into 7 different themes.
We provide an overview in Figure~\ref{fig:template-results-figure}.
The themes covered three topics: requirements, structure, and usability.
The spreadsheet \theme{requirements}{\faListUl} shaped the physical structure of the table as the requirements would inform the table schema.
Similar to \citet{huang2024table}'s study, we found the structure of the spreadsheet was hierarchical and was dived at the \theme{spreadsheet level}{\faNewspaper} and \theme{table level}{\faTable}.
We also elucidated new themes related to different types of tables that provided organization within the spreadsheet.
This included \theme{data tables}{\faDatabase} (see Figure~\ref{fig:template}-2), which contained the fundamental information about the data schema, and \theme{insight tables}{\faLightbulb[regular]} (see Figure~\ref{fig:template}-1), which used the data tables to generate insights.
These types of tables reflect how professionals create spreadsheets, which involves developing the underlying schema to represent the data and then generating insights based on the defined schema~\cite{pirolli2005sensemaking}.
Finally, the user experience of the spreadsheet, which included themes like \theme{ease of use}{\faThumbsUp[regular]} and \theme{presentation}{\faPalette}), enhanced the programmer's experience of using spreadsheets.
An example Excel spreadsheet template is shown in Figure~\ref{fig:template}.
We now discuss the codes for each theme in detail below.

\subsubsection{\faThumbsUp[regular]\xspace Ease of Use}
The Excel templates included features to make them usable by other programmers. 
This included \formativecode{documentation}, such as notes, tool tips, or guides on how to use the spreadsheet.
Written guides were commonly included as a separate sheet within the Excel workbook.
Templates were often explicitly denoted as \formativecode{simple} by including terms such as "basic" or "simple" in the title.

\subsubsection{\faPalette\xspace Presentation}
As noted in prior literature~\cite{huang2024table}, we identified a theme related to the presentation of the template.
This included the \formativecode{theme} of the template and \formativecode{conditional formatting} of the cell data.
We also noted \formativecode{interactive elements} of the spreadsheets, such as navigation elements.

\subsubsection{\faListUl\xspace Requirements}
One theme was the requirements for the spreadsheet, which was not discussed in prior work.
The templates differed according to \formativecode{audience} (e.g., for personal use versus business use) and \formativecode{timescale} (e.g., budgets on a monthly versus a yearly basis).
The templates also varied by additional \formativecode{context} about the purpose of the spreadsheet (e.g., a budget for a wedding, marketing team, family meal plan, or landscaping project).
The requirements influenced the spreadsheet table schema.
For example, a spreadsheet for a gardening project could require columns for the name, type, and quantity of each plant, while a spreadsheet for a yearly business budget could require columns for each month of the year.

\subsubsection{\faNewspaper\xspace Spreadsheet Level}
Template features occurred at the spreadsheet level, such as each sheet having a unique \formativecode{spreadsheet name}.
There were also differences in the overall \formativecode{architecture} of the spreadsheet.
Some spreadsheets put all the tables on a single sheet (e.g., in Figure~\ref{fig:template}), while others separated the data and insight tables in different sheets.

\subsubsection{\faTable\xspace Table Level}
Template features also occurred at the table level.
Each table was given a \formativecode{table name} and a \formativecode{schema} of its underlying data.
\formativecode{Example rows} and data were often provided to illustrate the schema and to show how the formulas worked.
For example, in Figure~\ref{fig:template}, there are example data in the data and insight tables that demonstrate the behavior of the aggregation in the last row, as well as the conditions under which an up or down arrow appears.
In addition, tables often included \formativecode{transforms}, which were cells that had formulas that encoded computational relationships between multiple columns in the schema.
For example, Figure~\ref{fig:template} includes the "Over / Under" column that takes the difference between the estimated and actual costs.

\subsubsection{\faDatabase\xspace Data Tables}
Data tables, which represented the underlying data for the spreadsheet, had unique features.
Two data tables are included in the example template in Figure~\ref{fig:template}.
Data within the data tables were either \formativecode{categorical values} (i.e., represented by categorical data) or \formativecode{literals} (i.e., plain values).
As noted in \citet{huang2024table}'s study, at the bottom of the data tables, there were often \formativecode{aggregation cells} that aggregated raw data (e.g., total cost for column A) to summarize the information in the table.
Finally, these tables had \formativecode{parameter cells} that contained related information that was not incorporated into the schema (e.g., a checkbox to track if the data corresponding to a table were paid).

\subsubsection{\faLightbulb[regular]\xspace Insight Tables}
Insight tables, designed to transform data tables into actionable insights, featured distinct elements from the data tables themselves. 
An example of an insight table can be found in the template shown in Figure~\ref{fig:template}. 
Complex spreadsheets with multiple tables often included an \formativecode{aggregation reference cell}, which directly linked to an aggregation cell within a data table. 
These could be further aggregated using a \formativecode{table aggregation cell} to provide an overall summary of the spreadsheet. 
For instance, in Figure~\ref{fig:template}, the insight table connects the cost of each category through aggregate reference cells, which are then summarized in the bottom row via a table aggregation cell. 
Additionally, insight tables were often enhanced with \formativecode{charts} to visually represent the data.

\mybox{
\faArrowCircleRight\xspace\textbf{Key findings (RQ1):}
The design space of spreadsheet templates include the structure and design of the spreadsheet.
The spreadsheet's design is influenced by the requirements of the problem and usability aspects, such as ease of use and presentation. 
In terms of spreadsheet structure, spreadsheet template features span at the spreadsheet level and table level.
Tables come in two variants: data tables that define the underlying data within the sheet and insight tables that summarize the data contained in the data tables.
}

\section{Formative User Study}
\label{sec:formative-user-study}
While \citet{erwig2006gencel}'s user study of Gencel provided some insights on scaffolding support through spreadsheet templates, it is unclear what the benefits and drawbacks of scaffolding tools are compared to having no guidance.
To support programmers in creating spreadsheets through scaffolding, we conducted a formative user study to understand current practices in using scaffolding tools in spreadsheet programming.
In particular, we wanted to understand programmers' experiences of using existing scaffolding tools (i.e., spreadsheet templates) and how this compared to not receiving any scaffolding assistance. 
We also wanted to understand spreadsheet programmers' needs for how an AI could assist them in creating a spreadsheet.
Therefore, the research questions in this study were:
\begin{description}
\setlength{\itemsep}{0pt}
\setlength{\parskip}{2pt}
    \item[\textbf{RQ2}] What is the experience of spreadsheet programmers while developing spreadsheets from scratch?
    \item[\textbf{RQ3}] What is the experience of spreadsheet programmers while developing spreadsheets from a template?
    \item[\textbf{RQ4}] What do spreadsheet programmers want from an AI that assists with creating spreadsheets?
\end{description}

To answer RQ2, RQ3, and RQ4, we recruited 7 spreadsheet programmers (Section~\ref{sec:formative-user-study-participants}) who completed a spreadsheet programming task (Section~\ref{sec:formative-user-study-task}) during an hour-long user study (Section~\ref{sec:formative-user-study-protocol}).
We then analyzed the data (Section~\ref{sec:formative-user-study-analysis}) to understand how programmers developed spreadsheets from scratch and from templates as well as their needs for an AI spreadsheet creation tool (Section~\ref{sec:formative-user-study-results}).
Using the findings from this study, we define design principles (Section~\ref{sec:design-goals}) for our tool for building spreadsheets, \tool.
The materials used in this study, including the protocols and codebooks, are included in the supplemental materials~\cite{supplemental-materials}.

\subsection{Participants}
\label{sec:formative-user-study-participants}
We recruited 7 active spreadsheet programmers through =UserTesting, an online user study platform. 
We ensured participants were familiar with spreadsheet software, as the study focused on usage rather than learning barriers.
Therefore, the inclusion criteria were Excel users who were working professionals from English-speaking countries and used Excel for at least 5 days in the previous month.
Our participants were men ($N=3$) and women ($N=4$) aged 26 to 49 ($\mu=35$) located in the United Kingdom ($N=2$), United States ($N=2$), and Canada ($N=3$).
Participants used Excel for 11-20 days ($N=4$) and 21+ days ($N=3$) in the previous month.
Participants reported experience with a variety of features in Excel.
Most participants had experience with basic Excel features, such as simple formulas like sum or average ($N=7$), formatting and layout ($N=5$), and copying data into Excel ($N=5$). 
Fewer participants reported experience with sorting and filtering ($N=2$) as well as pivot tables or pivot charts ($N=1$).

\subsection{Task}
\label{sec:formative-user-study-task}
We developed the user study task so that it included elements of the process that professionals follow to create knowledge products such as spreadsheets~\cite{pirolli2005sensemaking}: gathering information about the task, schematizing the task data, and developing insights based on the schema.

\paragraph{Task Description}
The user study task was in the domain of task tracking.
We selected this domain because similar to the budget domain in the spreadsheet template study (see Section~\ref{sec:template-analysis}), it was a predefined category of templates in the Excel application.
The user study task presented participants with a list of food drive tasks and asked them to determine who was doing the most work.
Tasks varied by frequency (e.g., twice a week versus once a quarter) and duration.

\subsection{Protocol}
\label{sec:formative-user-study-protocol}
The user study used a moderated observational study protocol followed by a debriefing interview, and was conducted remotely on a teleconferencing platform, where the participant's screen was recorded.
During the user study, the participants created spreadsheets to complete the task twice: once from scratch and once from a template retrieved from the task tracking template category in Excel version 2409.
This is because we wanted to observe participants creating spreadsheets with and without the assistance of scaffolding tools.
To reduce learning effects, the order of the conditions was roughly counterbalanced.

\paragraph{Overview}
To begin the study, the researcher introduced the user study task.
The participant then completed the task either by creating a table from scratch or from a template in Excel.
Participants were allowed to access the Internet or use any Excel feature to replicate a typical work environment.
Participants were initially given 15 minutes to complete the task and were provided extra time to edit the spreadsheet if needed.

Participants who created a spreadsheet from scratch were given a blank Excel workbook, while those who used a template received a copy of the task planning template. 
After completing the first task, participants repeated it under the opposite condition. 
Following each round, they debriefed about their experience and, after both rounds, compared and contrasted the two approaches.

At the end of the study, participants were shown three designs of an AI that created spreadsheets which that varied in the level of guidance provided by the tool.
The designs were presented as high-fidelity screenshot prototypes based on Excel and Excel Copilot. 
The first design involved interacting with the AI from the Excel canvas as a formula, while the second option used chat-based interaction. 
The most structured option featured an AI-driven setup wizard with static UI elements like drop-down menus. 
Participants were asked to share their impressions of these designs.

\paragraph{Piloting}
Following best practices for user studies in software engineering~\cite{ko2015practical}, we piloted the user study protocol with two spreadsheet programmers to: 1) ensure the quality of the study protocol by identifying and reducing potential confounding factors and 2) verify that the task allowed participants to follow the process detailed by \citet{pirolli2005sensemaking}.
We updated the wording of the protocol based on participants' feedback. 
We found that all pilot participants completed the task in the 15-minute time frame and engaged in gathering information about the task, schematizing the data, and developing insights.

\subsection{Analysis}
\label{sec:formative-user-study-analysis}
After the study session, one author performed transcription and two authors qualitatively analyzed the data, following the same qualitative analysis process in Section~\ref{sec:template-analysis-analysis}.
The authors developed an initial codebook on 2 transcripts (25\% of the samples).
Since the codes were not very skewed and there were two coders, we again used Cohen's $\kappa$~\cite{halpin2024inter}.
Following the same code masking process from Section~\ref{sec:template-analysis-analysis} with one  transcript, we obtained a Cohen's $\kappa$ score of 0.84---near perfect agreement~\cite{landis1977measurement}.
No new codes were discovered while applying the codes to the remaining data.

We also performed an analysis of the participants' development activities to study their development process. 
The second author reviewed footage of the spreadsheet development process and applied a codebook on programmer implementation actions from~\citet{liang2023qualitative} at the category level.
Codes included \emph{ideating} (e.g., brainstorming solutions), \emph{assessing} (e.g., reading or evaluating the current spreadsheet), \emph{prototyping} (e.g., testing different spreadsheet schema), \emph{implementation} (e.g., filling in table data, writing formulas), \emph{verifying} (e.g., reviewing the solution) and \emph{updating knowledge} (e.g., searching online for documentation).
Each action was labeled with a start time, a stop time, and a code label, resulting in a dataset of 217 actions.
To understand the effect of the task on the amount of time spent on each action, we aggregated the time each participant spent on every action.
We then ran a mixed-effects model that modeled participants as a random effect, with an interaction term between the conditions and actions: 

\begin{equation*}
\texttt{Total Action Time} \sim \texttt{Condition} + \texttt{Action} + \texttt{Condition} \times \texttt{Action} + (\texttt{1|Participant})
\end{equation*}

\begin{figure}[t!]
\centering
\includegraphics[trim=75 1650 75 0, clip, width=\linewidth, keepaspectratio]{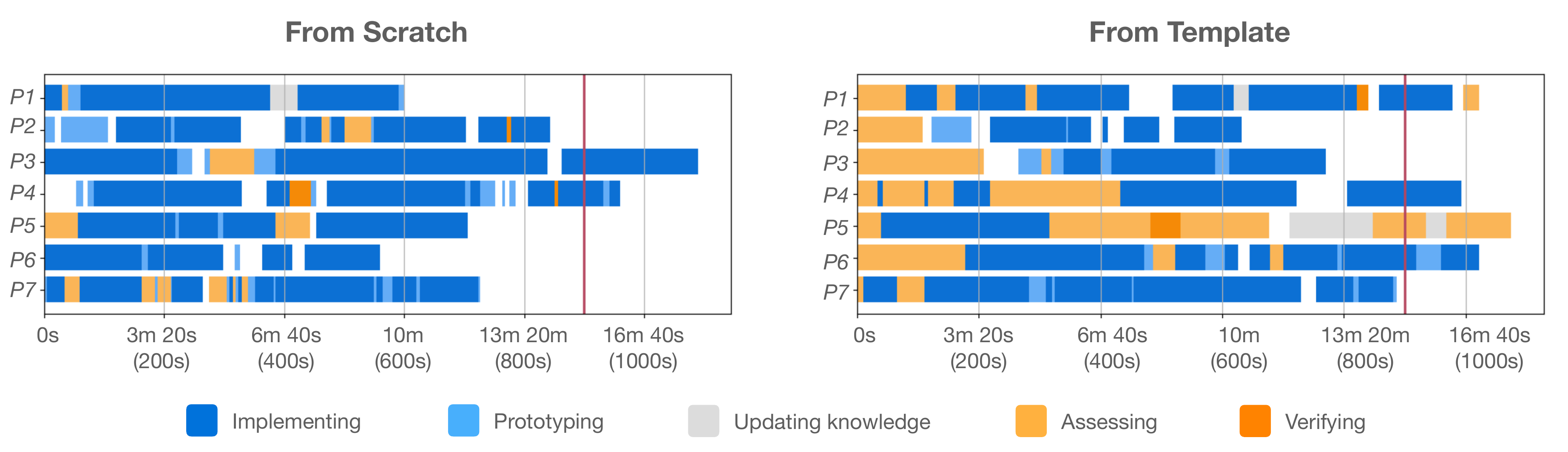} 
\caption{
A timeline of the participants' spreadsheet programming activities from the formative study for creating a spreadsheet from scratch (left) and from a template (right).
The red vertical line denotes the 15-minute mark.
}
\label{fig:formative-study-activity}
\end{figure}

\subsection{Results}
\label{sec:formative-user-study-results}
An overview of the activities of the participants is in Figure~\ref{fig:formative-study-activity}.
We find that creating spreadsheets from scratch and from templates both have benefits and drawbacks.
Creating from scratch affords flexibility and customization of the spreadsheets, but can reduce the overall quality of the final spreadsheet due to minimal guidance on performing Excel actions (e.g., writing formulas).
Meanwhile, starting from a template provides structure to the problem solving process and produces more polished spreadsheets. 
However, using a template requires significant adaptation that can interfere with the programmer's ability to complete the task.
We now discuss the results in further detail.

\subsubsection{From Scratch}
While creating spreadsheets from scratch, we found that 4 of the 7 participants completed the task in the 15 minutes allotted.
2 participants completed the task given extra time; only P5 did not complete the task.
The study participants spent most of their time on implementation.
Each participant followed their own unique process to implement a spreadsheet (see Figure~\ref{fig:formative-study-activity}), which aligns with traditional software engineering~\cite{liang2023qualitative}.
All participants completed multiple rounds of prototyping and implementation, supporting previous findings that the spreadsheet creation process is highly iterative~\cite{huang2024table, pirolli2005sensemaking}.
In addition, we noted that P2---the only participant who created a polished spreadsheet with formulas and themes---modeled a professional's process of first prototyping the table schema and then implementing the rest of the spreadsheet~\cite{pirolli2005sensemaking}.

Creating spreadsheets from scratch enabled \formativecode{customized} ($N=5$) spreadsheets that fit the programmer's context, as it was not constrained by a predefined structure.
This allowed the spreadsheet to be \pquote{[designed] in a way that naturally makes sense to me}{7}, making it \pquote{quite easy...to move stuff around}{6}.
However, the resulting sheets were \formativecode{quick-and-dirty} ($N=3$) that were not well-designed: \pquote{The table is not structured. It's more like a note thing rather than a table}{1}.
Participants ($N=5$) did not include any formatting elements (e.g., table themes or borders) in their spreadsheets.
Furthermore, starting from scratch provided \formativecode{limited support on structuring problems} ($N=7$).
Participants noted difficulties knowing \pquote{how...to present [information] in the most natural way}{7} and \pquote{putting my thought process into the spreadsheets}{4}.
This echos previous work that indicates that knowing how to schematize data is a skill to develop spreadsheets~\cite{pirolli2005sensemaking, ko2011state, chalhoub2022s} and that problem-solving knowledge is an important skill in programming~\cite{baltes2018towards, liang2022understanding}.
Finally, participants had \formativecode{minimal guidance on using advanced Excel features} ($N=5$), especially formulas: \pquote{I'm not too familiar with the formulas... That's why it took me a bit longer to get the calculations}{3}.
Some participants attempted computations with Excel formulas, but gave up due to difficulties using the feature and instead developed workarounds.
P1 resorted to using their computer's calculator, while P3 performed arithmetic mentally.
This corroborates prior findings that programmers struggle with understanding spreadsheet programming APIs~\cite{ko2011state, nardi1991twinkling}.
Not using spreadsheet formulas could reduce the accuracy of the resulting calculations; for example, P3 resorted to rounding numbers and estimating results.

\mybox{
\faArrowCircleRight\xspace\textbf{Key findings (RQ2):} 
Programmers who develop spreadsheets from scratch create spreadsheets that match the programmer's context, but the resulting spreadsheets require additional effort from the programmer to become a polished product}.
Programmers struggle to structure the table to solve the task and lacked guidance on using advanced Excel features to implement the spreadsheet.
In addition, a vast majority of programmers completed the task by creating spreadsheets from scratch.

\subsubsection{From a Template}
Only a single participant (P7) completed the task in the 15 minutes allotted when using an Excel template to create a spreadsheet.
Although 4 of the 7 participants required additional time, these participants were still unable to complete the task.
This suggests that the usage of templates hindered the completion of the task compared to creating from scratch.
This aligns with previous work, which finds that the use of programming tools can increase task completion time~\cite{pandita2018no} and increase task failure rates~\cite{vaithilingam2022expectation}.

In addition, we observed that participants spent additional time on all actions compared to creating spreadsheets from scratch, with the increase in time of each action ranging from 1.4 minutes (83 seconds) to 4.0 minutes (237 seconds).
The largest differences in time spent occurred in assessing the spreadsheet ($p=0.003$) and updating knowledge ($p=0.40$), where study participants spent an additional 4.0 minutes (237 seconds) understanding the template and 2.1 minutes (123 seconds) looking up documentation.
In terms of process (see Figure~\ref{fig:formative-study-activity}), several participants ($N=4$) entered multiple rounds of implementation and prototyping, indicating that spreadsheet development is iterative~\cite{huang2024table} even while working with templates.
However, fewer participants performed these iterations compared to creating spreadsheets from scratch, suggesting that some participants adopted the spreadsheet template schema with minimal modification of the columns.

Participants noted that templates were \formativecode{structured} ($N=4$), which could help spreadsheet programmers with problem-solving: \pquote{These templates are set up to be quick and easy if you can’t think of how you want to do something}{6}.
However, study participants found that creating spreadsheets from templates was difficult for a variety of reasons.
For example, the structure imposed by the template was too rigid and \formativecode{required adaptation} ($N=7$) to fit the programmer's context, such as removing extraneous columns: \pquote{Templates [are] quite annoying because they're not necessarily catered 100\% to my needs}{6}.
This has been observed in prior studies on programming scaffolding tools, which find that the tool needs to be adapted to the programmer's context~\cite{erwig2006gencel, latoza2020explicit}.
Because spreadsheet templates required significant effort for \pquote{modification than it is actually worth}{7}, templates were useful only when they closely matched the problem: \pquote{The template...feels very specific.  You can only use it for work that is exactly the same as the one here}{3}.
In addition, templates were often \formativecode{hard to understand} ($N=5$), as noted in prior work~\cite{reschenhofer2015empirical}: \pquote{I'm not familiar enough...to interpret at a glance what someone else has created}{5}.
Since the participants \pquote{did not set up [the spreadsheet] myself}{5} and struggled to understand \pquote{how calculations works}{1}, they \pquote{need[ed] to spend more time...working out if the template is designed for me}{6}.
However, participants noted several benefits of templates.
This included offering \formativecode{reusable features} ($N=3$) like drop-down lists for validation, filtering, or formulas.
Participants also noted templates were more \formativecode{polished} ($N=6$): \pquote{I do like the visual presentation and the layout of the template}{3}.

\mybox{
\faArrowCircleRight\xspace\textbf{Key findings (RQ3):} 
Spreadsheet programmers who develop spreadsheets from templates have more structure to solve the problem and produce more polished spreadsheets. 
However, the templates require heavy adaptation and are difficult to understand.
Finally, creating spreadsheets from templates requires more time to assess the templates, which can result in less task completion.
}

\subsubsection{Spreadsheet Programmer Needs for AI}
The majority of the participants preferred the chat bot ($N=6$) compared to the setup wizard ($N=1$) and the in-canvas formula ($N=0$), since the chat bot was more \formativecode{flexible} ($N=7$) and could be used to \pquote{update and rectify [outputs]}{1} via natural language.
The participants felt that the setup wizard interface was \pquote{too structured and too limited}{2} since a programmer would be dependent on \pquote{how extensive the options are}{5}, whereas for the in-canvas interface, there was no \pquote{interaction capability}{1}.

The participants wanted to \formativecode{describe tasks and goals} ($N=5$) to the AI, since their \pquote{concerns are always more related to the high level}{7}: \pquote{You don't want to say, 'I want cell A1 + B12`... You want to explain the scenario}{1}.
Next, participants said the AI should have \formativecode{understandable outputs} ($N=3$) for easy modification: \pquote{If [the AI] was not transparent about what it created, I'd have a hard time...making little tweaks}{5}.
Further, participants said the tool should be \formativecode{usable} ($N=6$) and not require technical knowledge: \pquote{I have a decent base of coding, but I want to explain what I want without using any technical terms}{1}.
Participants also noted the tool should provide suggestions on \formativecode{insights} ($N=5$) in the data, like specific analyses, \pquote{graphs}{3}, or \pquote{ways to summarize the data}{5}.

\mybox{
\faArrowCircleRight\xspace\textbf{Key findings (RQ4):}
For an AI that assists with creating spreadsheets, spreadsheet programmers prefer a chat-based interface for its flexibility in expressing intent and correcting outputs.
Programmers want to work with the AI to express high-level intents and understand the AI's outputs for a seamless modification of the spreadsheet.
}

\section{Design Principles}
\label{sec:design-goals}
Based on the findings from the spreadsheet template study (Section~\ref{sec:template-analysis}) and the formative user study (Section~\ref{sec:formative-user-study}), we reason that an effective tool for developing spreadsheets should follow these design principles (labeled "DP"):

\paragraph{\textbf{\dgA}}
Tools that support spreadsheet programming should guide users through the process of creating spreadsheets.
Spreadsheet programmers struggle with \formativecode{limited support on structuring problems} and having \formativecode{minimal guidance on using advanced Excel features} (e.g., writing formulas) while creating spreadsheets from scratch, which results in \formativecode{quick-and-dirty} spreadsheets.
In contrast, templates offer welcome \formativecode{structure}.

Prior literature shows that scaffolding tools can improve programmer task performance in a variety of activities~\cite{latoza2020explicit, arab2022exploratory, bajpai2024lets}. 
We thus infer that guiding spreadsheet programmers through established processes can achieve their goals more effectively, especially since professional spreadsheet programmers follow a defined process of gathering information, representing information in a schema, developing insight on the data, and creating a knowledge product~\cite{pirolli2005sensemaking}.
For spreadsheet programming, whereby the process inherently creates a knowledge product (i.e., a spreadsheet), spreadsheet programming tools should guide users through the process of: 1) gathering \theme{requirements}{\faListUl} about the spreadsheet table, 2) defining the data schema in \theme{data tables}{\faDatabase}, and 3) extracting insights by creating \theme{insight tables}{\faLightbulb[regular]}.
For gathering requirements, the tool should elicit the \formativecode{audience} and additional \formativecode{context} of the spreadsheet such as \formativecode{timescale}.
To create data tables, the tool should create a \formativecode{schema} with the given requirements and generate \formativecode{example rows}.
To develop insight tables, the tool should suggest and generate analyses to obtain insights from the data.

\paragraph{\textbf{\dgB}}
Tools supporting spreadsheet programming should imbue flexibility within the structured process since scaffolding tools can impose too much rigidity~\cite{reiser2018scaffolding, arab2022exploratory}.
They should have \formativecode{flexible} interactions because spreadsheet development is a highly iterative and non-linear process and spreadsheet programmers follow their own unique implementation process (see Section~\ref{sec:formative-user-study-results}).
However, spreadsheet programming scaffolding tools are inflexible and \formativecode{require adaptation} to become \formativecode{customized} to fit the \theme{requirements}{\faListUl} of the programmer.
Therefore, we reason that it is important for tools that support spreadsheet programming to provide flexibility to perform actions that fit a programmer's specific context for better task performance.

\paragraph{\textbf{\dgC}}
Tools that support spreadsheet programming should build spreadsheets iteratively and incrementally.
A drawback to creating spreadsheets from a template is that templates are \formativecode{hard to understand} due to their complexity, requiring programmers to spend more time understanding the template.
Therefore, tools that support spreadsheet programming should produce \formativecode{understandable outputs} that are \formativecode{simple}.
We infer that incrementally building spreadsheets with programmers in smaller units can result in less manual adaptation of the generated spreadsheets and make the spreadsheet creation easier for the programmer.

\begin{figure}[t!]
\centering
\includegraphics[trim=0 175 825 0, clip, width=0.95\linewidth, keepaspectratio]{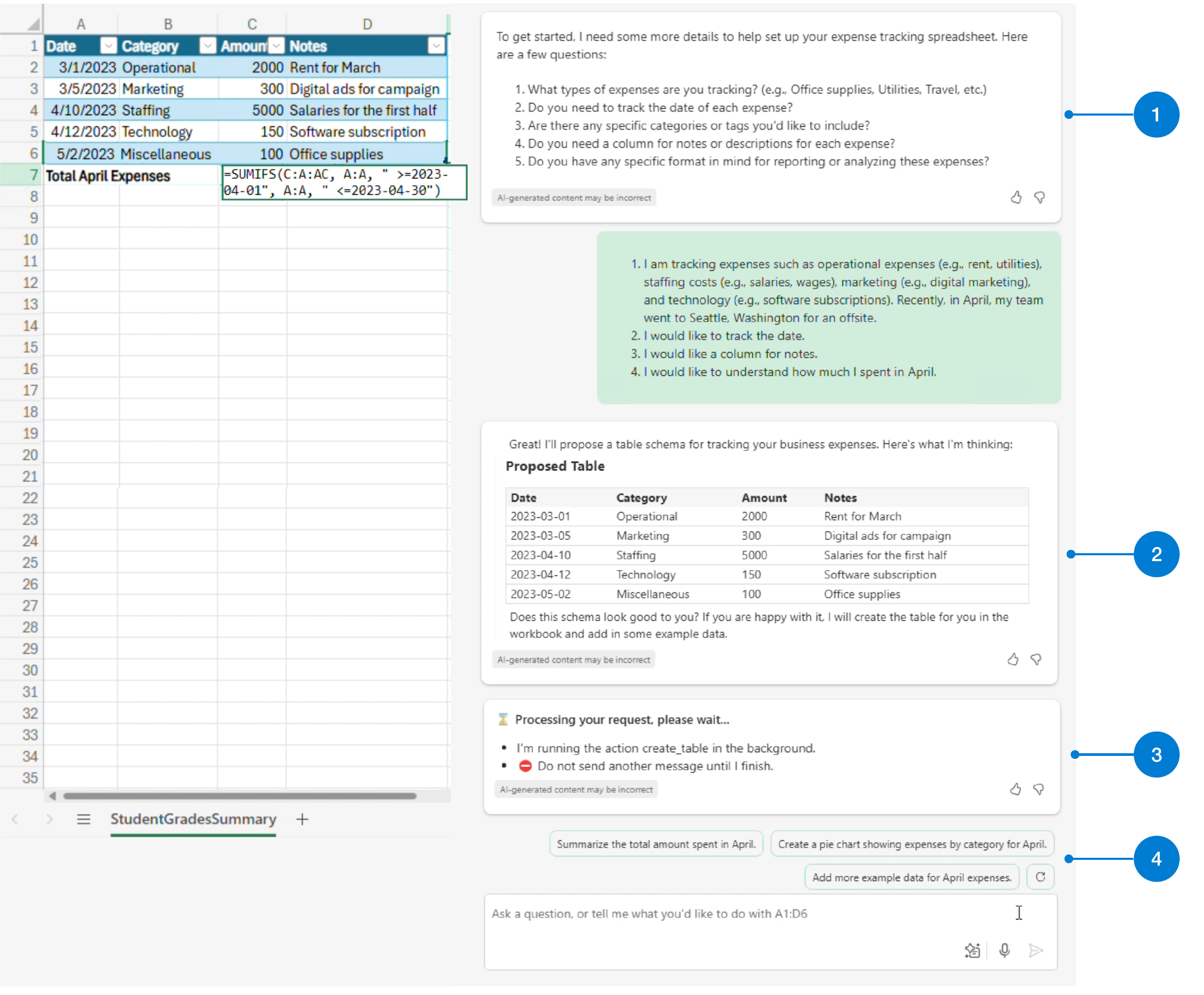} 
\caption{
The interface of \tool, an agent-based tool that helps programmers build spreadsheets in Excel by \ilabel{1} guiding programmers through a structured plan of developing spreadsheets (gathering requirements, defining the data schema, and extracting insights) based on expert processes; \ilabel{2} prototyping tables in the chat using Markdown; \ilabel{3} leveraging tools that produce spreadsheet components can be composed to build spreadsheets; and \ilabel{4} suggesting next steps based on the plan and the current state to adapt the plan to the programmer's context.
}
\label{fig:tool}
\end{figure}

\begin{table*}
  \centering
\caption{
A comparison of the state-of-the-art research and state-of-the-practice spreadsheet creation tools based on the design principles (DP; Section~\ref{sec:design-goals}) and features (F; Section~\ref{sec:design}).
\tool is the only option that provides scaffolding support (DP1) that is flexible (DP2) for spreadsheet creation through incremental development (DP3).
}
\label{tab:tool-comparison}
\begin{tabular}{p{0.4\linewidth}|cc|cc}
\toprule
& \multicolumn{2}{c}{\textbf{Research}} & \multicolumn{2}{c}{\textbf{Practice}} \\
\textbf{Design Principle \& Feature} & \textbf{\tool} & \textbf{SheetCopilot~\cite{li2024sheetcopilot}} & \textbf{Excel Copilot} & \textbf{Templates}\\ 
\hline
\rowcolor[rgb]{ .921,  .921, .921}
\multicolumn{5}{l}{\textbf{\dgA}}\\
\hline
\fA & \multirow{2}{*}{\faCheck\xspace Yes} &  \multirow{2}{*}{\faTimes\xspace No} &  \multirow{2}{*}{\faTimes\xspace No} &  \multirow{2}{*}{\faTimes\xspace No} \\
\hline
\rowcolor[rgb]{ .921,  .921, .921}
\multicolumn{5}{l}{\textbf{\dgB}}\\
\hline
\fB & \multirow{2}{*}{\faCheck\xspace Yes} &  \multirow{2}{*}{\faTimes\xspace No} &  \multirow{2}{*}{\faTimes\xspace No} &  \multirow{2}{*}{\faTimes\xspace No} \\
\midrule
\fC & \multirow{2}{*}{\faCheck\xspace Yes} & \multirow{2}{*}{\faTimes\xspace No} & \multirow{2}{*}{\faCheck\xspace Yes} & \multirow{2}{*}{\faTimes\xspace No} \\
\hline
\rowcolor[rgb]{ .921,  .921, .921}
\multicolumn{5}{l}{\textbf{\dgC}}\\
\hline
\fD & \multirow{2}{*}{\faCheck\xspace Yes} & \multirow{2}{*}{\faCheck\xspace Yes} & \multirow{2}{*}{\faCheck\xspace Yes} & \multirow{2}{*}{\faTimes\xspace No} \\
\bottomrule
\end{tabular}
\end{table*}

\section{\tool System}
\label{sec:tool}
We implement \tool, a tool that assists programmers in creating spreadsheet programs in Excel (see Figure~\ref{fig:tool}).
Its main novelty lies not in its interface---which is a standard chat---but in its implementation of the design principles for a spreadsheet programming experience that is scaffolded but flexible, and creates spreadsheets incrementally, rather than all at once.

We compare \tool with other state-of-the-art spreadsheet programming tools in research (i.e., SheetCopilot~\cite{li2024sheetcopilot}) and practice (i.e., Excel Copilot, spreadsheet templates) based on the design principles (labeled "DP") and features (labeled "F") elaborated below (see Table~\ref{tab:tool-comparison}).
\tool is the only tool that achieves all design principles.
For example, templates provide scaffolding, but do not offer flexibility or enable incremental spreadsheet development.
Agentic tools such as SheetCopilot~\cite{li2024sheetcopilot} and Excel Copilot can perform spreadsheet development actions, but do not provide scaffolding for the programmer.
We note that Excel Copilot offers Markdown table previews in the chat and while it offers suggestions, the suggestions are not based on the spreadsheet creation plan.
Now, we introduce \tool in a motivating example (Section~\ref{sec:motivating-example}) and then discuss the design of \tool (Section~\ref{sec:design}) and its implementation details (Section~\ref{sec:implementation}).

\subsection{Motivating Example}
\label{sec:motivating-example}
We revisit Alma from Section~\ref{sec:introduction}, who wants to analyze her small business expenses to understand her top spending categories for her business co-owner.
Due to Alma's earlier struggles to implement this spreadsheet, she tries \tool:

\scenario{
Alma opens Excel and launches \tool's chat pane, where it asks about her spreadsheet goals. 
She replies, "I need a spreadsheet to track spending categories for my small business." 
\tool guides Alma through the requirements by asking clarifying questions about task details (Figure~\ref{fig:tool}-1). 
Alma responds, "It’s for me and my co-owner to identify top spending categories. Here’s a recent credit card statement..." and pastes the statement into the chat.
\newline\newline
\tool prototypes a sample spreadsheet as a Markdown table in the chat (Figure~\ref{fig:tool}-2), with columns for date (A), category (B), amount (C), and notes (D), populated using Alma’s credit card data. 
Although Alma could continue to edit the table in the chat, she is now satisfied with the schema, and she tells \tool to transfer the table to Excel.
In response, \tool runs the action to create a table (see Figure~\ref{fig:tool}-3), adding it to the Excel canvas.
\newline\newline
With the table now added to Excel, \tool presents three potential next steps that are displayed as rounded, clickable UI elements known as "suggestion pills": \emph{"Summarize the total amount spent in April."}, \emph{"Create a pie chart showing expenses by category for April."}, and \emph{"Add more example data for April expenses."} (see Figure~\ref{fig:tool}-4).
Alma selects the first option as it best aligns with her goal. 
The tool prototypes a table with two columns---category and total cost. 
The total cost column includes an Excel formula, such as \texttt{=SUMIFS(C:C, A:A, ">=2023-04-01", A:A, "<=2023-04-30"}, B:B, "Operational"), to calculate costs by category for the past month. 
Alma accepts, and the table is added below the first one and sorted by total cost.
Next, Alma selects the suggestion \emph{"Create a pie chart showing expenses by category for April."}, which generates a pie chart using the new table’s data. 
Satisfied with the results, she shares the spreadsheet with her co-owner for review.
}

\subsection{Design}
\label{sec:design}
To reify the design principles in \tool, we leverage language agents since they can achieve the design goals of scaffolding, flexibility, and incrementality. 
This is because LLMs can autonomously reason through complex tasks and can adapt to feedback in their environment~\cite{shinn2024reflexion, gupta2024metareflection, paul2023refiner, madaan2024self}, which can provide both scaffolding and flexibility.
By leveraging tools (i.e., external modules that complete specific actions) to build atomic components of spreadsheets, agents can construct spreadsheets incrementally.
We further discuss the main features of \tool and how they support the design principles below.

\subsubsection{\dgA}
Generative AI, like language agents, can be effective metacognitive scaffolds for programmers~\cite{prather2023s, leinonen2023comparing, franklin2025generative, xing2025use}. 
We therefore designed \tool to scaffold programmers through the structured process that professionals follow to develop knowledge products from data, such as spreadsheets (\emph{F1}).

\paragraph{\fA} 
Following prior LLM tools that guide users through expert processes~\cite{sun2024reviewflow, bajpai2024lets}, \tool scaffolds programmers through the spreadsheet creation process enumerated in Section~\ref{sec:design-goals}.
We anticipate that following a plan could alleviate the metacognitive burden on spreadsheet programmers by minimizing the need to verbalize task decomposition, which often occurs while interacting with LLMs~\cite{tankelevitch2024metacognitive}.
We translated each step as a natural language instruction and included these instructions in the tool's system prompt as a step-by-step recipe (see Section~\ref{sec:implementation}).
For the first step of gathering requirements, the agent elicits the programmer's spreadsheet needs by asking questions about the audience, context, and timescale of the spreadsheet as well as requesting related documents.
For the second step of creating a data table, the agent proposes a schema with the given information and generates a table with example rows.
Finally, to develop insight tables, \tool suggests different insights that a programmer could obtain from the data and implements an analysis that the programmer requested.

\subsubsection{\dgB}
\tool is designed to offer flexibility in the structured process, ensuring programmers are not constrained by rigid scaffolding. 
The chat interface allows programmers to freely interact with the tool and express their needs without limitations. 
Additionally, suggestion pills (\emph{F2}) reveal potential next steps, enabling users to adapt the process based on their personal context for human-in-the-loop planning. 
The tool also supports rapid prototyping of spreadsheet tables in Markdown (\emph{F3}), providing programmers with an efficient way to iterate and refine their ideas.

\paragraph{\fB}
LLMs can perform better decision-making by considering multiple reasoning paths and looking ahead to the next step~\cite{yao2024tree}.
\tool provides multiple suggestions to help programmers explore the solution space, leveraging their problem-solving expertise~\cite{baltes2018towards, liang2022understanding} to select the most suitable action.
This approach ensures the language agent performs actions that align closely with the programmer's goals.
Similar to prior work using LLMs to generate follow-up suggestions~\cite{bajpai2024lets}, \tool employs LLM-generated follow-ups during each interaction.
However, unlike previous systems, \tool's follow-up responses are designed to guide programmers through the expert process the agent is following, advancing their progress in a structured yet adaptable manner.
To guide the programmer through the agent's expert process, \tool provides three next-step suggestions displayed as suggestion pills. 
These are generated based on the conversation history, spreadsheet context, and the programmer's goal, ensuring the most efficient path to achieving the objective.

\paragraph{\fC}
Prototyping is vital in software development to evaluate solution viability~\cite{liang2022understanding}. 
Given that spreadsheet development is iterative~\cite{pirolli2005sensemaking, huang2024table}, rapid spreadsheet prototyping is crucial for testing schemas. 
\tool enables this through Markdown because it is human-readable and is a widely supported format that LLMs can generate~\cite{liu2024we}.
While direct manipulation in Excel would be intuitive, it is technically impractical due to the slow inference speed of language agents, which requires planning and executing actions~\cite{wang2024survey}. 
Instead, \tool uses chat-based prototyping, offering faster iterations and the ability to test solutions without modifying the spreadsheet since text generation is more efficient than generating and performing actions.
In the future, improved agent reasoning speeds could make direct manipulation a feasible and desirable option.

\begin{table*}
  \centering
\caption{
A list of the tools \tool can use while developing spreadsheets with a user.
}
\label{tab:tools}
\begin{tabular}{p{0.2\linewidth}p{0.75\linewidth}}
\toprule
\textbf{Tool} & \textbf{Description}\\
\hline
\rowcolor[rgb]{ .921,  .921, .921}
\multicolumn{2}{l}{\theme{Spreadsheet level}{\faNewspaper}} \\
\hline
\texttt{change\_sheet\_name} & Updates the sheet name to the new sheet name. \\
\hline
\rowcolor[rgb]{ .921,  .921, .921}
\multicolumn{2}{l}{\theme{Table level}{\faTable} / \theme{Data table}{\faDatabase}} \\
\hline
\texttt{create\_table} & Creates a table within the sheet with the given name and a list of values to include in the table.\\
\hline
\rowcolor[rgb]{ .921,  .921, .921}
\multicolumn{2}{l}{\theme{Insight table}{\faLightbulb[regular]} } \\
\hline
\texttt{add\_chart} & Creates a chart (line, pie, histogram) for the column in a table (for numeric data only). \\
\texttt{sort\_rows} & Sorts the table rows based on the values in the given column. \\
\texttt{filter\_rows} &  Filters the table to rows that match the given condition. This does not permanently delete rows from the tables. \\
\hline
\rowcolor[rgb]{ .921,  .921, .921}
\multicolumn{2}{l}{\theme{Presentation}{\faPalette}} \\
\hline
\texttt{highlight\_cell} & Highlights the cell in the color (red, green, or yellow) that match the given condition. \\
\texttt{highlight\_row} & Highlights the entire row in the color (red, green, or yellow) if any value in the row matches the condition. \\
\texttt{change\_table\_color} & Changes the color of the given table. \\
\bottomrule
\end{tabular}
\end{table*}

\subsubsection{\dgC}
\label{sec:dgC}
\tool supports the incremental development of spreadsheets to produce changes that are easy for spreadsheet programmers to understand. 
We do so by creating tools that the language agent can use to implement the atomic components of spreadsheets (\emph{F4}).
The output of these tools can be combined to implement more complex spreadsheet program behaviors.

\paragraph{\fD}
Because language agents struggle to perform certain tasks (e.g., mathematical reasoning~\cite{lu2023survey}), language agents are commonly equipped with "tools" like calculators and code interpreters for additional capabilities~\cite{xi2023rise, wang2024survey, schick2024toolformer, gao2023pal}.
These tools function like API calls, where the agent populates parameters to execute tasks (e.g., running code with a code interpreter).
\tool uses tools to implement atomic spreadsheet components (e.g., conditional formatting, charts) by leveraging OfficeScript, an Excel-specific scripting language with APIs for direct canvas manipulation~\cite{microsoft2024officescript}. 
A tool-based approach is required since state-of-the-art LLMs like \texttt{GPT-4} have limited OfficeScript generation capabilities~\cite{payan2023instructexcel}. 
In \tool, pre-written OfficeScript scripts handle tasks, with the agent supplying parameters to execute desired behaviors and Excel executing these scripts to update the spreadsheet.
\tool tools include creating tables, changing table colors, highlighting rows and cells, visualizations, sorting, filtering, and renaming sheets. 
These tools, summarized in Table~\ref{tab:tools}, were designed to cover each spreadsheet structure-related theme from in the template study (see Section~\ref{sec:template-analysis}) for comprehensive component coverage.

\subsection{Implementation}
\label{sec:implementation}
\tool adapts the front end of Excel Copilot in Excel version 2409. 
It has a custom Flask server as the back end which contains the language agent implemented using the OpenAI Assistants API~\cite{openai-api-assistants} with \gpt.
We describe different implementation details of \tool, including the architecture (Section~\ref{sec:tool-architecture}), as well as the agents and their prompt structures (Section~\ref{sec:tool-agents}) below.

\begin{figure}[t!]
\centering
\includegraphics[trim=0 250 275 0, clip, width=\linewidth, keepaspectratio]{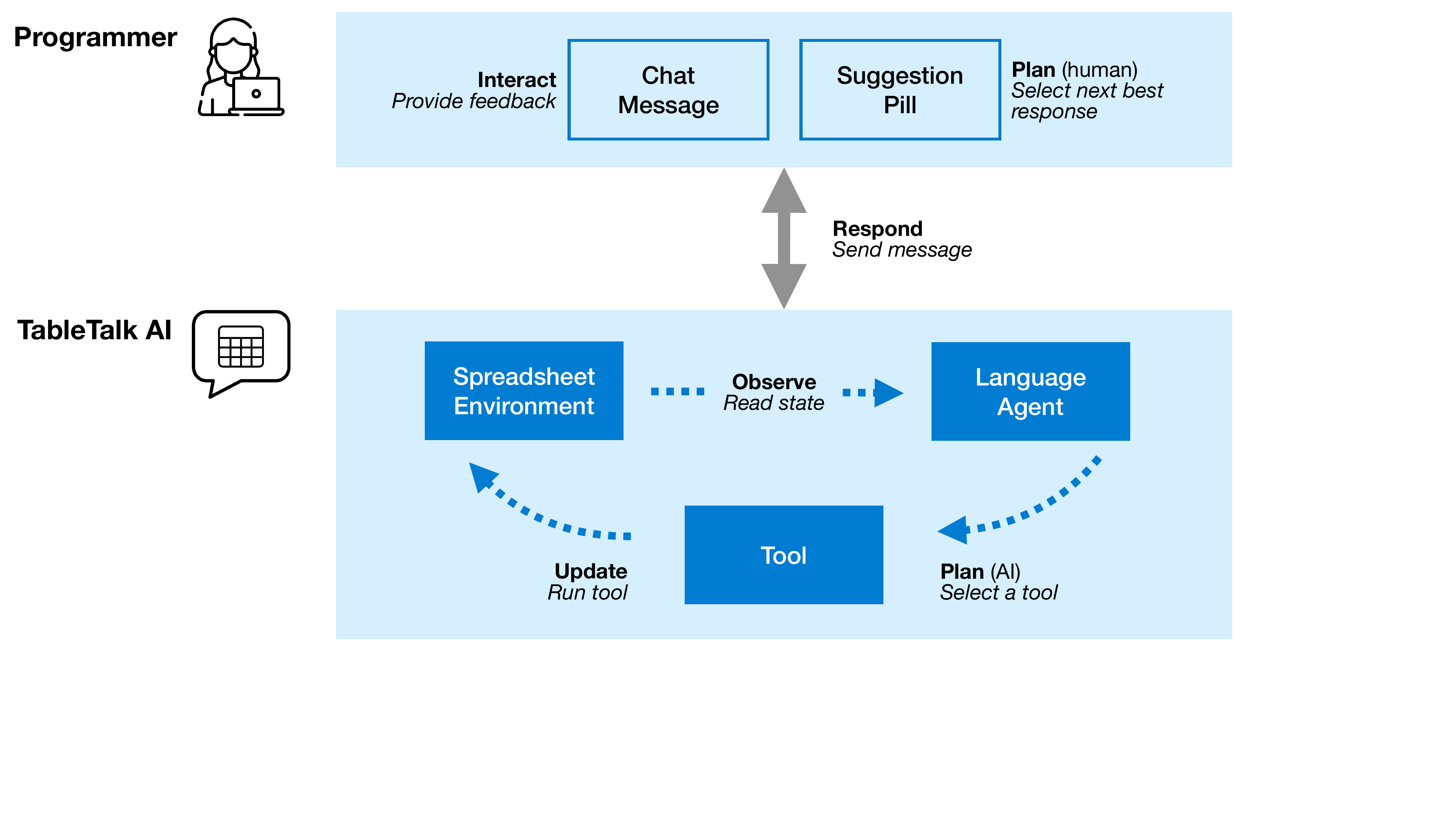} 
\caption{
An overview of \tool. 
\textbf{Interact/Observe:} A spreadsheet programmer interacts with \tool and \tool observes the current state of the spreadsheet. 
\textbf{Plan (AI):} Based on this information, \tool plans which tool to use to address the programmer's query. \
\textbf{Update:} After selecting a tool, \tool executes the tool to update the spreadsheet. 
\textbf{Observe:} The result of running the tool, including the new workbook state, is sent back to \tool. 
\textbf{Respond:} \tool responds with a summary of the changes that it made. 
\textbf{Plan (human):} The spreadsheet programmer assists with planning by selecting the next step through the suggestion pill. 
\tool reads the new spreadsheet state, and decides its next step. 
At any given interaction, \tool can continue to execute tools (\textbf{Observe}, \textbf{Plan (AI)}, \textbf{Update}) or respond to the programmer (\textbf{Respond}).
}
\label{fig:overview}
\end{figure}

\subsubsection{Architecture}
\label{sec:tool-architecture}
The front end gathers input, captures current workbook state, executes tools and records the result of the execution, and finally renders the back end's response to the programmer.
Meanwhile, the back end's responsibility is to run the language agent. 
The agent considers the programmer's query and the current workbook state, generates a response to the programmer, and selects tools that will complete the programmer's request.

Figure~\ref{fig:overview} illustrates the interaction between the front end and back end. 
When a user sends a message, the front end generates a textual representation of the workbook's current state and sends it, along with the user's message, to the back end. 
The back end uses the OpenAI Assistants API to determine the best tool for the query and returns a response with tool calls, which are executed in the front end.
The updated workbook state is saved and sent back to the back end to keep the API informed. If additional tools are required, the back end requests further executions, and this loop continues until no tools remain to run.

\subsubsection{Agents}
\label{sec:tool-agents}
\tool is implemented as a single agent system.
It operates in two steps: one to generate utterances and plan actions based on the current state and another to generate suggested follow-up responses to the programmer.

The agent analyzes the current spreadsheet state and chat history to generate an utterance or select appropriate tools. 
If an OfficeScript tool is chosen, the agent populates its parameters according to the tool's specification, which includes details on its behavior, parameter descriptions, and expected output. 
Each parameter is validated to ensure it is well-formed. If an error is detected, the agent receives an error message and regenerates the parameter.
Once all parameters are error-free, the tool is sent to the front end for execution.
If an utterance is generated, the agent generates three suggestions for potential next steps in a separate API call.
The agent considers the current spreadsheet state, chat history, the most recent message, and the goals of the programmer and generates three suggestions for the programmer.
For this prompt, we adopt the ReAct prompting approach~\cite{yao2023react}, where the model simulates the programmer's thought process at each step before producing a suggestion. 
This encourages deeper reasoning by the model, leading to more contextually relevant suggestions.

The agent's prompt begins with a description of the Excel spreadsheet programming context.
It then outlines a JSON representation of each table, structured as a JSON array containing metadata for each cell (e.g., cell address, value, formula). 
Next, the prompt details the step-by-step spreadsheet creation process, including instructions for prototyping tables in Markdown when defining data and insight tables. 
The agent is also tasked with summarizing the programmer's current overall goal. 
Finally, additional guidelines are provided, such as using Excel formulas and ensuring constraints like non-overlapping tables.
Meanwhile, the suggestion prompt begins with a description of the Excel spreadsheet programming context and the overall task.
Guidelines for writing the suggestions are provided (e.g., being concise or suggesting actions for only existing tables) and finally, the JSON output format of the suggestion is described.

\section{Evaluation}
\label{sec:evaluation-user-study}
To evaluate \tool's agentic approach to implementing the design principles, we recruited 20 spreadsheet programmers (Section~\ref{sec:evaluation-user-study-participants}) to perform two tasks (Section~\ref{sec:evaluation-user-study-tasks}) in a controlled study (Section~\ref{sec:evaluation-user-study-protocol}) using \tool and a \baseline.
The baseline for this study is the Excel Copilot available in Excel version 2409 (see Table~\ref{tab:tool-comparison}).
We then analyzed these data quantitatively and qualitatively (Section~\ref{sec:evaluation-user-study-analysis}) to better understand the results of our research questions (Section~\ref{sec:evaluation-user-study-results}).
The materials used in this study, including the protocols and codebooks, are included in the supplemental materials~\cite{supplemental-materials}.

One objective of this study was to derive lasting insights into human-agent collaboration that would remain relevant over time. 
While current agentic approaches like \tool~\cite{wang2024survey} are characterized by high latency, we anticipate future technological advancements will reduce these delays. 
To address this, our evaluation focused on spreadsheet quality rather than task completion, with study protocols and analyses designed to consider latency effects.
Therefore, our research questions for the evaluation are as follows:

\begin{description}
\setlength{\itemsep}{0pt}
\setlength{\parskip}{2pt}
    \item[\textbf{RQ5}] How does using \tool affect the quality of the created spreadsheets?
    \item[\textbf{RQ6}] How does using \tool affect the way a spreadsheet programmer creates spreadsheets?
    \item[\textbf{RQ7}] How does using \tool affect a spreadsheet programmer's perceived experience in creating spreadsheets?
\end{description}

\subsection{Participants}
\label{sec:evaluation-user-study-participants}
We recruited 20 spreadsheet programmers via an online user study recruitment platform known as PlaybookUX.
We applied the same inclusion criteria as the formative user study (see Section~\ref{sec:formative-user-study-participants}).
Our participants were men ($N=12$) and women ($N=8$) aged 28 to 60 ($\mu=40$) located in the United Kingdom ($N=4$), United States ($N=9$), Canada ($N=6$), and Australia ($N=1$).
Participants used Excel 5-10 days ($N=3$), 11-20 days ($N=7$) or 21+ days ($N=10$) in the previous month.
Like the formative study, almost all participants had experience with basic Excel features, such as simple formulas like sum
or average ($N=19$), formatting and layout ($N=17$), copying data into Excel ($N=16$), and sorting and filtering ($N=15$).
Some participants reported experience with more advanced features, such as pivot tables or pivot charts ($N=11$), macros ($N=6$), and Power Query or data connections to another workbook ($N=7$).

\subsection{Tasks}
\label{sec:evaluation-user-study-tasks}
Participants completed two tasks: 1) analyzing student grades and 2) tracking hours worked and client payments. 
These tasks were selected from \citet{chalhoub2022s}'s study of spreadsheet programmers and aligned with common occupations requiring spreadsheet skills~\cite{glass2024skills}. 
Each task had its own reference solution with example data.
The written task instructions directed participants to create spreadsheets that were polished enough for reuse by others and were visually appealing to encourage high-quality table creation. 
To promote ecological validity, the written task instructions included only the problem statement, without specific data. 
This reflects real-world spreadsheet programming, where users often start with broad problem definitions and refine the data as they work~\cite{pirolli2005sensemaking}.

Like the formative study task (see Section~\ref{sec:formative-user-study-task}), these tasks were designed to reflect the professional process of creating knowledge products~\cite{pirolli2005sensemaking}. 
Since spreadsheet programmers struggle to schematize and implement tables~\cite{pirolli2005sensemaking, ko2011state, chalhoub2022s}, we designed the task to be difficult in problem solving (i.e., knowing how to approach the problem) and implementation (i.e., writing Excel formulas), which we reason that AI tools such as \tool and the \baseline could assist with.
The tasks were intentionally difficult to complete within the 15-minute time limit to focus on studying variance in spreadsheet quality rather than obtaining high-quality spreadsheets.
Participants were expected to make substantial progress but not necessarily finish the tasks, which we confirmed through piloting (see Section~\ref{sec:evaluation-user-study-protocol}).

\paragraph{Task Descriptions}
The student grades analysis task required participants to count the overall grades of all students in a course based on their raw grades for various classroom activities. 
To prevent \tool and the \baseline from automatically using a standard classification scale (e.g., A–F), the overall grades were on a custom scale of Excellent ($\geq$ 95\%), Very good ($\geq$ 85\%), Satisfactory ($\geq$ 75\%) and Needs improvement ($<$ 75\%).
The overall grade for each student was calculated as a weighted average of their performance across classroom activities, which were divided into assignments, participation, and exams. 
Each category had a unique weight and a varying number of activities.

The task of tracking hours worked and client payments required participants to calculate the total hours worked in September based on a list of tasks completed for various clients. 
Tasks spanned multiple months, and each task type (e.g., consulting) had an associated hourly rate. 
For hours exceeding 10 hours, an overtime rate of 10\% above the base hourly rate applied. 
For instance, 11 hours of work at \$100 per hour would result in the first 10 hours charged at \$100 each and the 11th hour charged at \$110.

\subsection{Protocol}
\label{sec:evaluation-user-study-protocol}
The user study was a controlled and moderated observational study with a debriefing interview and surveys.
It was 90 minutes long and was conducted remotely on a teleconferencing platform.
To reduce bias, the two tools were framed as different versions of Excel Copilot, given that \tool was adapted from the baseline.
To reduce the confounding factors and learning effects associated with having two tools and two tasks, the order of the tasks and the tools was counterbalanced in four conditions.
Each participant was assigned to a single condition, where the participant used two tools for two different tasks.
This resulted in each condition being associated with 5 data points. 

\paragraph{Overview}
At the start of the study, participants were introduced to the study context, which was framed as a spreadsheet programmer creating a spreadsheet for a client. 
Then, the researcher shared their screen and gave the participant control of the researcher's computer to ensure access to the tool.

The participant then completed a 10-minute warm-up activity in which the participant created a simple budget spreadsheet.
The goal of the warm-up activity was to teach the participant the tool's capabilities and how to interact with the tool via open-ended text and suggestion pills.
In addition, for the \baseline, we instructed participants to include a table with headers in the spreadsheet for the best results.
During the tutorial, the participant developed a preliminary spreadsheet with data and then completed four sub-tasks: changing the color of the table, highlighting the cells, generating graphs, and sorting the table.
After completing the tutorial, the participant was briefed on the task and given time to read the specific task instructions in a separate document.
After answering any questions about the task, the researcher provided additional instructions.
To promote the use of the AI tool, the participant was instructed to first try performing an action with the AI and then manually if the tool failed.
The participant was also encouraged to clarify the spreadsheet requirements with the researcher.

Participants were then given 15 minutes to complete each task which was enforced with a timer. 
Halfway through the task, the researcher provided the participant with an example row from the reference solution to allow the participant to evaluate their spreadsheet and ensure consistency in the spreadsheet.
At the end, the participant completed a questionnaire to collect data on perceived workload, system usability, and conversation quality.
The questionnaire contained the NASA Task Load Index (TLX)~\cite{hart1988development} instrument on a 10-point scale, System Usability Scale (SUS)~\cite{brooke1996sus} instrument on a 5-point scale, and two questions on conversation quality.
These questions were related to the tool's \emph{proactivity} and \emph{relevance} in conversation, which are the strongest predictors of overall conversation quality~\cite{finch-choi-2020-towards}.
After completing the questionnaire, the warm-up exercise and task protocol were repeated for the next task and tool. 

After completing both tasks with \tool and the baseline, the researcher conducted a semi-structured interview where the participant answered questions about the benefits and drawbacks of each tool and selected which tool they preferred overall.
To account for the latency of \tool, if the participant mentioned \tool's response time, the researcher asked whether the participant's decision would change if the response times between the tools were equal.

\paragraph{Piloting}
We piloted the study protocol with four spreadsheet programmers, following the same process as the formative user study (see Section~\ref{sec:formative-user-study-protocol}). 
The pilot confirmed that the tasks allowed participants to follow the professional process of working with open-ended data~\cite{pirolli2005sensemaking}.
Participants made significant progress within the 15-minute time limit, despite not completing the tasks due to their designed difficulty. 
All participants successfully created well-formed spreadsheets for the task using at least one of the tools.

\subsection{Analysis}
\label{sec:evaluation-user-study-analysis}
We performed quantitative and qualitative analyses on the questionnaire data, transcripts, session recordings, and participant spreadsheets.

\paragraph{Spreadsheet Analysis}
To assess the quality of participant spreadsheets (RQ5), we conducted a convenience sampling of six spreadsheet programmers who had performed spreadsheet programming in the past month. 
Each evaluator was assigned a task and was provided with the corresponding instructions.
Evaluators compared 10 pairs of spreadsheets, with each pair consisting of one created using \tool and one from the baseline. 
The order of the pairs was randomized to prevent evaluators from knowing which tool was used for each spreadsheet.
For each pair, the evaluator selected the spreadsheet they preferred to reuse for continuing the task and provided open-ended feedback explaining their choice.
Each task had 3 evaluators (30 comparisons total), with each pairing being unique across each task.

We quantitatively analyzed these data by reporting descriptive statistics on evaluator preferences and modeling the strength of \tool- and \baseline-generated spreadsheets using a Bradley-Terry model.
Bradley-Terry statistical models can estimate an ability score (i.e., relative strength) of teams in competitions using pairwise comparisons~\cite{agresti2012categorical}.
In our Bradley-Terry model, each team corresponds to a tool (\tool or the baseline) and the resulting ability score of \tool represents the log-likelihood that \tool~is preferred to the baseline.

We triangulated our quantitative results with a qualitative analysis of the 60 written responses from spreadsheet evaluators. 
Due to the simplicity of the data, only the first author developed the codebook. 
This was done by extracting all comments about participant spreadsheets from the open-ended responses, yielding 166 total comments. 
These comments were open-coded based on the aspects of the spreadsheets they addressed, with each assigned a single code and labeled as either "positive" or "negative."
To validate the codebook, we used Cohen's $\kappa$ due to having two coders---one developing the codebook and one validating it---and a relatively balanced distribution of codes~\cite{halpin2024inter}.
The second author applied the codebook to a random 25\% subset of the data using the same code masking procedure in prior studies.
This process resulted in near-perfect agreement~\cite{landis1977measurement}, with a Cohen's $\kappa$ score of 0.88. 
To assess whether \tool influenced the evaluators' comments, we conducted a Pearson's chi-square test on the counts of the positive and negative codes between the two tools.
To understand which codes contributed to this difference, we examined the residuals of the chi-square test and identified those whose value exceeded $\pm2$~\cite{sharpe2015your}.

\paragraph{Task Analysis}
To understand how participants created spreadsheets (RQ6), we analyzed the study's video recordings. 
An author reviewed the recordings and applied \citet{mozannar2024reading}'s CodeRec User Programming States (CUPS) taxonomy, which includes codes for a tool's response time. 
Since the taxonomy was designed for traditional programming contexts, we added a new code for manually entering data, a common activity in spreadsheet programming.
Each activity was labeled with its start time, stop time, and a code, resulting in a dataset of 976 actions. 
To determine if there were differences between the tools, we conducted a Pearson's chi-square test on the activity code frequencies and examined the standard residuals. 
Additionally, we used a mixed-effects modeling approach, including an interaction term between each tool and action, following the analysis approach from the formative user study (Section~\ref{sec:formative-user-study-analysis}).
\begin{equation*}
\texttt{Total Action Time} \sim \texttt{Tool} + \texttt{Action} + \texttt{Tool} \times \texttt{Action} + (\texttt{1|Participant})
\end{equation*}

We also conducted a qualitative analysis of the chat logs. 
After each study session, the first author extracted chat messages from each tool, resulting in 257 messages. 
Using the same single-author coding and validation procedure as in the spreadsheet analysis, we masked the codes of 25\% of the samples and used Cohen's $\kappa$ for IRR due to having two coders and a relatively even distribution of codes~\cite{halpin2024inter}.
The codebook validation achieved a Cohen's $\kappa$ score of 0.80, indicating substantial agreement~\cite{landis1977measurement}.
During the analysis, we identified six non-substantive confirmation messages (e.g., "Yes" or "This looks good") that were excluded due to their lack of qualitative insight, allowing the analysis to focus on substantive interactions. 
To explore differences in the types of messages sent between the two tools, we performed a Pearson's chi-square test on the message codes and examined the standard residuals.

\paragraph{Semi-Structured Interview Analysis}
To understand participants' experiences creating spreadsheets (RQ7), we transcribed their responses to the semi-structured interview questions about their tool usage. 
Two authors conducted a qualitative analysis of the transcripts, following the same procedure used in the template study (see Section~\ref{sec:template-analysis-analysis}) and the formative user study (see Section~\ref{sec:formative-user-study-analysis}).
To validate the codes, we repeated the code masking procedure to one transcript and computed IRR, using Cohen's $\kappa$ since we had similar conditions as the prior analyses.
The IRR of the codebook using Cohen's $\kappa$ was 0.79, indicating substantial agreement~\cite{landis1977measurement}. 
The first author found no new codes after applying the codebook to the remaining data.

\paragraph{Post-Task Questionnaire Analysis}
To study participants' experiences creating spreadsheets (RQ7), we conducted a quantitative analysis of the post-task questionnaire data. 
We calculated an aggregated SUS score using the standard SUS scoring procedure~\cite{lewis2018system}.
We then used Wilcoxon signed-rank tests to compare the aggregated SUS scores, individual TLX items, and individual conversational quality items between \tool and the baseline. 
For constructs with five or more comparisons, we applied Benjamini-Hochberg corrections, which affected only the \emph{p}-values of the NASA TLX items.

\begin{table*}
  \centering
\caption{
An overview of the comments evaluators made about the spreadsheets made using \tool and the \baseline.
We provide a description of each code and the number of times the code appeared for each tool ($N=n$).
The percentages in italics ($NN\%$) in the chart represent the percent of the comments related to that code for the each tool that was a positive (left, green) and negative (right, red).
}
\label{tab:spreadsheet-results}
\begin{tabular}{p{0.1\linewidth}p{0.35\linewidth}p{0.47\linewidth}}
\toprule
\textbf{Code} &\textbf{Description} & \textbf{Distribution} \\ 
\midrule
\evalcode{polish} & \multirow{2}{\linewidth}{The spreadsheet is overall more polished or complete, such as having more formatting elements, data entered, or other extra helpful spreadsheet features.} & \mylabel{\tool ($N=28$)} \newline \polaritybarchart{0.86}{0.14}{0}{86\%}{14\%} \\
&  & \polaritybarchart{0.25}{0.75}{0}{25\%}{75\%} \newline \mylabel{Excel Copilot ($N=24$)} \\
\midrule
\evalcode{schema} &\multirow{2}{\linewidth}{The spreadsheet's columns (e.g. column names) or column data types are well-suited to address the problem.} & \mylabel{\tool ($N=18$)} \newline \polaritybarchart{0.56}{0.44}{0}{56\%}{44\%}\\
&  & \polaritybarchart{0.4}{0.6}{0}{40\%}{60\%} \newline \mylabel{Excel Copilot ($N=15$)} \\
\midrule
\evalcode{correctness} & \multirow{2}{\linewidth}{The spreadsheet's functionality (especially formulas) is generally correct.} & \mylabel{\tool ($N=28$)} \newline \polaritybarchart{0.5}{0.5}{0}{50\%}{50\%}\\
&  &  \polaritybarchart{0.37}{0.63}{0}{37\%}{63\%} \newline \mylabel{Excel Copilot ($N=19$)} \\
\midrule
\evalcode{usability} & \multirow{2}{\linewidth}{The spreadsheet is easy to use (e.g., having formulas versus plain values), simple, or easy to understand.} & \mylabel{\tool ($N=27$)} \newline \polaritybarchart{0.56}{0.44}{0}{56\%}{44\%}\\
&  & \polaritybarchart{0.29}{0.71}{0}{29\%}{71\%} \newline \mylabel{Excel Copilot ($N=14$)} \\
\midrule
\multicolumn{3}{c}{\mylegend{Positive}{green2} \mylegend{Negative}{red2}} \\
\bottomrule
\end{tabular}
\end{table*}

\subsection{Results}
\label{sec:evaluation-user-study-results}
During the study, no participant completed any task, as anticipated due to the difficulty of the task. 
However, P1, P4, and P18 progressed to summarizing the total counts of grades in the gradebook task using \tool ($N=2$) and the \baseline ($N=1$).
We now discuss the results on how \tool affected the quality of the created spreadsheets (RQ5; Section~\ref{sec:evaluation-user-study-results-spreadsheets}), how spreadsheet programmers created spreadsheets (RQ6; Section~\ref{sec:evaluation-user-study-results-actions}), and the experience of spreadsheet programmers (RQ7; Section~\ref{sec:evaluation-user-study-results-experience}).

\subsubsection{Spreadsheet Quality}
\label{sec:evaluation-user-study-results-spreadsheets}
Spreadsheet evaluators expressed a variety of comments on the generated spreadsheets (see Table~\ref{tab:spreadsheet-results}). 
We observed a difference in attitudes towards the spreadsheets created by \tool compared to the \baseline ($\chi^2=25.4$, $p<0.001$).
We further elaborate on these findings below.

\paragraph{\textbf{\tool produces higher quality spreadsheets compared to the baseline.}}
Across both tasks, spreadsheet evaluators preferred the spreadsheets created by \tool over the baseline 42 out of 60 times (70\%).
The general preferences of the individual evaluator towards \tool ranged from 50\% to 90\%.
Based on the Bradley-Terry model, \tool had an ability score of $0.85$, indicating that \tool's spreadsheets had 2.3 times higher odds of being preferred over the baseline's ($p<0.001$).

Evaluators expressed a marked difference in the spreadsheet \evalcode{polish}, a general-purpose code for the overall completeness of the spreadsheet.
Evaluators were much less negative ($r=-2.4$) for \tool ($p=0.02$) and much more negative ($r=2.8$) for the baseline ($p=0.005$) than expected.
We find support for this in the qualitative data.
While both tools received similar positive feedback, such as the spreadsheets having \equote{nicer formatting}{1} or being \equote{filled out}{1, E3, E5} and \equote{complete}{4, E5, E6},
the evaluators noted a distinct lack of polish for the baseline's spreadsheets, as some participants \equote{couldn't get anywhere with the task}{4}. 

\paragraph{\textbf{The \baseline's spreadsheets were rated negatively on polish, schema, correctness, and usability, while \tool's spreadsheets had positive or mixed ratings.}}
For \tool, the evaluator comments were strongly positive for \evalcode{polish} (86\%) and more mixed for \evalcode{usability} (56\%), \evalcode{schema} (56\%), and \evalcode{correctness} (50\%).
However, for the baseline, the comments trended strongly negative for \evalcode{polish} (75\%) and \evalcode{usability} (71\%) and slightly negative for \evalcode{schema} (60\%) and \evalcode{correctness} (63\%).

In terms of spreadsheet \evalcode{correctness}, evaluators praised spreadsheets from both tools for having correct formulas, but noted several instances when the spreadsheets did not have functionality associated with the end of the task, likely due to its difficulty. 
However, the spreadsheet evaluators noted that the \baseline's spreadsheets were \equote{completely wrong}{4}, such as not accounting for weighted averages.
In contrast, evaluators noted that the \tool spreadsheets also contained errors that were easy to fix ($N=4$) such as \equote{chang[ing] the range}{4}.

Spreadsheet evaluators discussed the \evalcode{usability} of the spreadsheets. %
They described instances where the spreadsheets of both tools were \equote{transparent}{3} on how the calculations were derived.
However, there were fewer comments ($N=4$, $r=-1.27$) of the baseline's spreadsheets being usable ($p=0.20$).
While the spreadsheets from both tools contained unexplained columns or values, the baseline's spreadsheets often contained hard-coded values ($N=4$) or complicated formulas, causing evaluators to struggle to understand them ($N=2$).

In terms of how the spreadsheet represented the information about the \evalcode{schema}, the evaluators had mixed comments for \tool (56\%) and slightly negative comments for the baseline (60\%).
Both tools received positive comments on \equote{the overall columns mak[ing] sense}{6} and negative comments on creating spreadsheets where the \equote{task requirements are not there}{6}. 
Yet, the evaluators said only the baseline created tables where the units did not make sense ($N=3$).

\mybox{
\faArrowCircleRight\xspace\textbf{Key findings (RQ5):}
\tool affects the quality of spreadsheets by producing better and more polished spreadsheets compared to the \baseline, leading to a strong preference of spreadsheets made by \tool.
Although the accuracy of \tool received mixed feedback, many of the errors could easily be fixed.
In contrast, the feedback on the baseline's tables was generally negative on their polish, schema, correctness, and usability.
}

\begin{table*}
  \centering
\caption{
An overview of the user messages sent to \tool and the \baseline.
}
\label{tab:chat-results}
\begin{tabular}{p{0.17\linewidth}p{0.35\linewidth}p{0.40\linewidth}}
\toprule
\textbf{Code} &\textbf{Description} & \textbf{Example Quote} \\ 
\midrule
\evalcode{requirements} & The message describes requirements of the problem (e.g., example data). & \pquote{Create a table to track the grades...there are 4 homework assignments and two exams and a participation activity}{1} \\
\midrule
\evalcode{high-level command} & A request to perform a spreadsheet action using language that does not reference specific parts of the spreadsheet. & \pquote{Almost there, can you also add a column for the date? Sept month needed only}{8} \\
\midrule
\evalcode{low-level command} & A request to perform a spreadsheet action by referencing specific parts of the spreadsheet (e.g., existing columns or cells) or describing computations. & \pquote{Insert Rate in front of column E}{8} \\
\midrule
\evalcode{help-seeking} & A request to gather information on how to perform an action or to explain a formula. & \pquote{How to calculate the final grade?}{10} \\
\bottomrule
\end{tabular}
\end{table*}

\begin{figure}[t!]
\centering
\includegraphics[trim=0 1000 0 0, clip, width=0.8\linewidth, keepaspectratio]{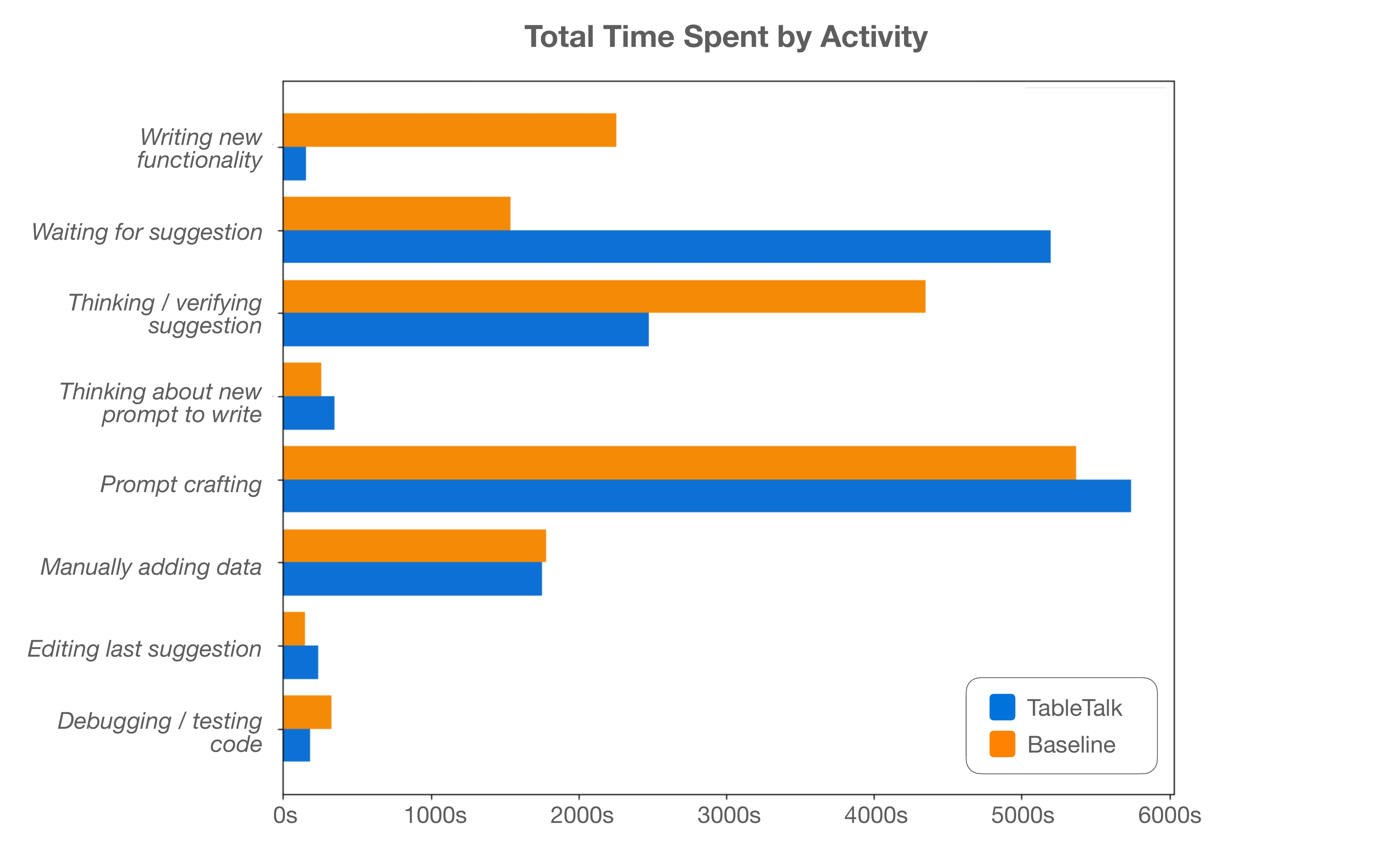} 
\caption{
The total number of seconds spent on different CodeRec User Programming States taxonomy~\cite{mozannar2024reading}  actions across all programmers for \tool (orange) and the \baseline (blue).
}
\label{fig:evaluation-study-activity-overview}
\end{figure}

\begin{figure}[t!]
\centering
\includegraphics[trim=225 700 225 25, clip, width=0.9\linewidth, keepaspectratio]{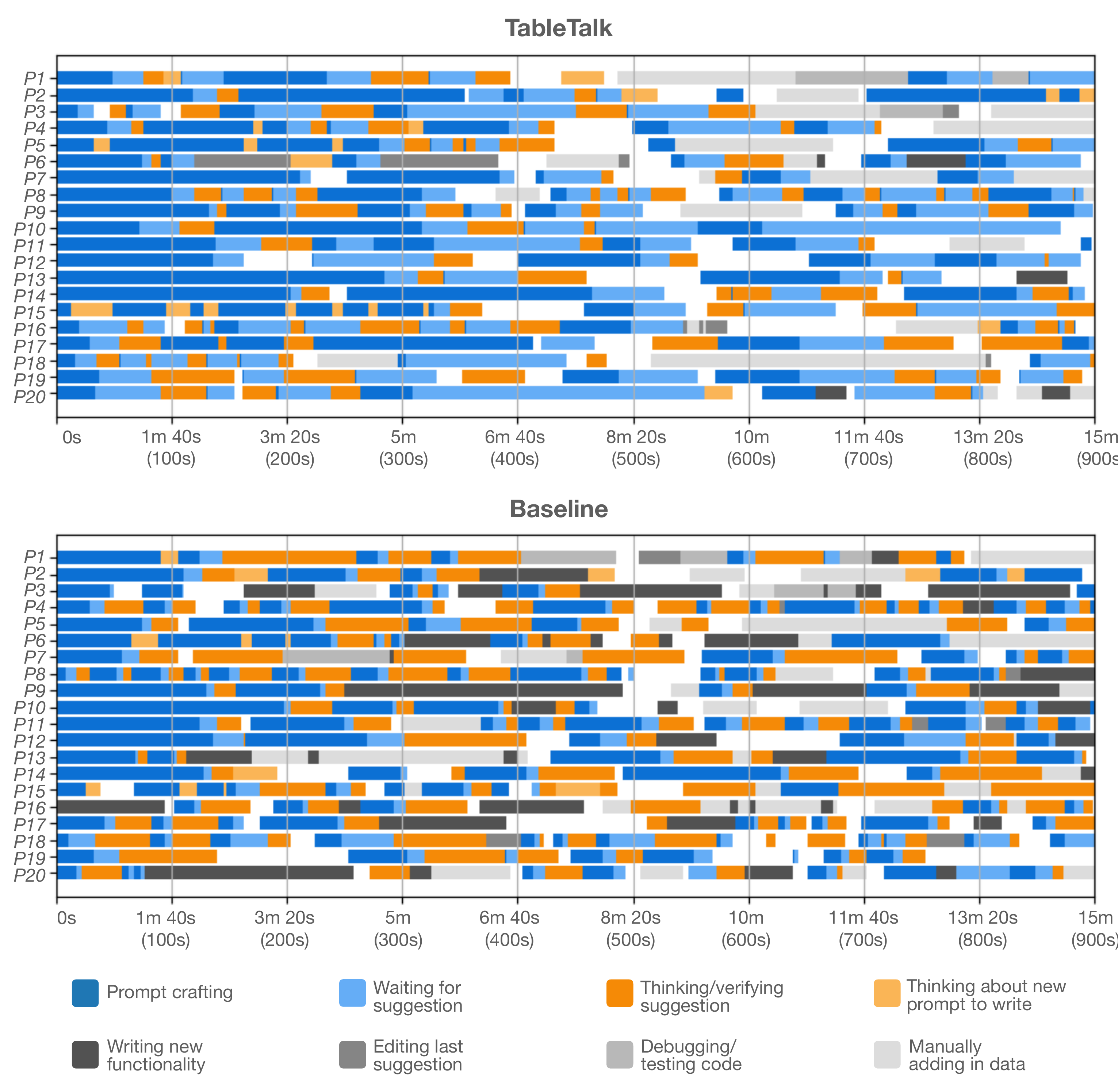} 
\caption{
A timeline of the participants' activities based on the CodeRec User Programming States (CUPS)~\cite{mozannar2024reading} for \tool (top) and for the \baseline (bottom).
}
\label{fig:evaluation-study-activity}
\end{figure}

\subsubsection{Spreadsheet Programmer Activity}
\label{sec:evaluation-user-study-results-actions}
We provide an overview of the time spent in each programming state based on the CUPS taxonomy (Figure~\ref{fig:evaluation-study-activity-overview}) and a summary of chat messages (Table~\ref{tab:chat-results}). 
Participants manually sent 131 chat messages to the baseline and 82 to \tool.
As expected, \tool programmers collectively spent 61.0 more minutes (3,657 seconds) waiting for suggestions compared to the baseline, averaging an additional 2.7 minutes (164 seconds) of waiting per participant ($p<0.001$). 
We also observed significant differences in overall activities ($\chi^2=31.1$, $p<0.001$) and the types of chat messages sent ($\chi^2=21.0$, $p<0.001$) between the tools.
We elaborate on these differences below.

\begin{figure}[t!]
\centering
\includegraphics[trim=20 1550 25 20, clip, width=0.90\linewidth, keepaspectratio]{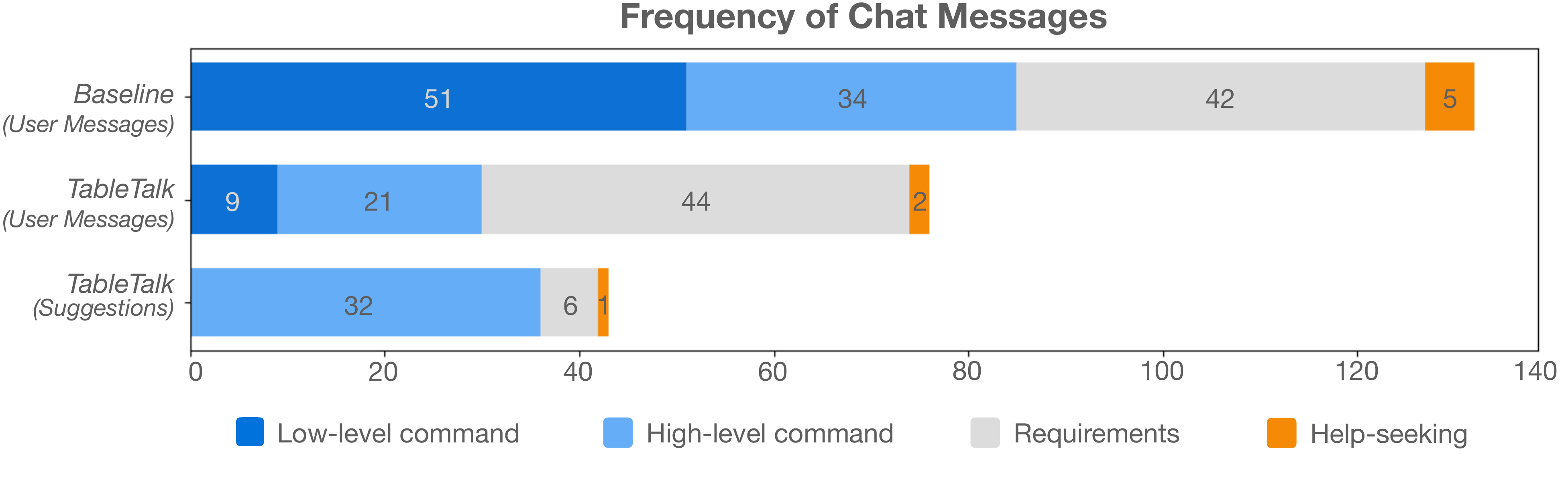} 
\caption{
The frequency of the different types of chat messages manually sent by participants between \tool and the \baseline.
The frequency of the chat message type is displayed within its bar.
}
\label{fig:evaluation-study-chats}
\end{figure}

\paragraph{\textbf{Programmers using \tool focus more on the requirements of the problem compared to the baseline.}}
Based on Figure~\ref{fig:evaluation-study-chats}, we observed a much higher presence of \evalcode{requirements} messages ($r=2.2$) sent to \tool ($p=0.03$) and a smaller amount ($r=-1.7$) of these messages sent to the baseline ($p=0.09$) than expected.
Examining the CUPS activity (Figure~\ref{fig:evaluation-study-activity}), participants spent longer continuous periods crafting their prompts, even though the total time spent prompting was similar across the two tools. 
This indicates that focusing on spreadsheet requirements can be labor-intensive, requiring sustained bursts of effort.

\paragraph{\textbf{Programmers using the baseline focus more on problem solving than compared to programmers using \tool.}}
Participants collectively spent an additional 31.2 minutes (1,871 seconds) thinking or verifying suggestions while using the baseline.
This translated into a 1.9 minute (112 seconds) reduction in thinking about spreadsheet programming actions or verifying the code while using \tool relative to the baseline ($p=0.007$).
In addition, 32 of the 39 accepted suggestions from \tool were \evalcode{high-level commands} (see Figure~\ref{fig:evaluation-study-chats}), indicating that the suggestions offloaded some of the problem solving from the programmer.

\paragraph{\textbf{Programmers using the baseline focus more on implementation details than compared to programmers using \tool.}}
While using the baseline, programmers spent an additional 35.0 minutes (2,100 seconds) writing new functionality across all participants.
Study participants wrote more new functionality ($r=3.0$) with the baseline ($p=0.003$) and less ($r=-3.4$) with \tool ($p<0.001$) than expected, resulting in an additional 2.0 minutes (118 seconds) of implementation while using the baseline ($p=0.07$).
In addition, the programmers sent significantly more \evalcode{low-level command} messages ($r=2.8$) using the baseline ($p=0.006$) and less ($r=-2.1$) for \tool ($p=0.04$) than expected (see Figure~\ref{fig:evaluation-study-chats}).
Finally, based on Figure~\ref{fig:evaluation-study-activity}, the participants went through multiple rounds of iterations to create prompts and implement table functionality while working with the baseline.
Although study participants spent less continuous time crafting prompts, they spent longer periods of time writing new functionality, indicating that working with \baseline required more intensive manual effort.

\mybox{
\faArrowCircleRight\xspace\textbf{Key findings (RQ6):} 
\tool affects how programmers creates spreadsheets by allowing programmers to focus more on the requirements of the spreadsheet instead of problem solving and implementation.
Programmers using \tool overall spend less time thinking about next steps and verifying suggestions, but spend more time waiting for suggestions.
They also write less new functionality.
Messages sent to \tool are less related to low-level commands and more related to requirements, while the suggestions are high-level commands.
}

\begin{table*}
  \centering
\caption{
An overview of the conversational quality~\cite{finch-choi-2020-towards} and NASA Task Load Index (TLX) results~\cite{hart1988development}.
NASA TLX results have a Benjamini-Hochberg correction applied.
Statistically significant ($p<0.05$) results are denoted with an asterisk (\raisebox{1pt}{\textnormal{\fontsize{5}{10}\selectfont\faStarOfLife}}).
The mean score of the distribution is denoted with $\mu$.
}
\label{tab:survey-results}
\begin{tabular}{p{0.45\linewidth}p{0.47\linewidth}}
\toprule
\textbf{Description} & \textbf{Distribution} \\ 
\hline
\rowcolor[rgb]{ .921,  .921, .921}
\multicolumn{2}{l}{
\textbf{Conversational Quality~\cite{finch-choi-2020-towards}}} \\
\hline
\multirow{2}{\linewidth}{The responses from the tool actively and appropriately moved the conversation along different topics.} & \mylabel{\tool ($\mu=3.8 / 5$)} \newline \cqbarchart{0.35}{0.35}{0.15}{0.05}{0.10}{0}{70\%}{15\%} \\
& \cqbarchart{0.2}{0.2}{0.25}{0.2}{0.15}{0}{40\%}{35\%}  \newline \mylabel{Baseline ($\mu=3.1 / 5$)} \\
\midrule
\multirow{2}{\linewidth}{\raisebox{1pt}{\textnormal{\fontsize{5}{10}\selectfont\faStarOfLife}} The responses from the tool were on-topic with the immediate dialogue history.} & \mylabel{\tool ($\mu=4.4/5$)} \newline \cqbarchart{0.55}{0.35}{0.05}{0.05}{0.0}{0}{90\%}{05\%} \\
& \cqbarchart{0.2}{0.3}{0.25}{0.2}{0.05}{0}{50\%}{25\%}  \newline \mylabel{Baseline ($\mu=3.4/5$)} \\
\midrule
\multicolumn{2}{c}{\mylegend{1 - Strongly disagree}{red3} \mylegend{2\xspace\xspace\xspace}{red2} \mylegend{3\xspace\xspace\xspace}{gray1} \mylegend{4\xspace\xspace\xspace}{green2}\mylegend{5 - Strongly agree}{green3}} \\
\hline
\rowcolor[rgb]{ .921,  .921, .921}
\multicolumn{2}{l}{
\textbf{NASA Task Load Index (TLX)~\cite{hart1988development}}} \\
\hline
\multirow{2}{\linewidth}{\raisebox{1pt}{\textnormal{\fontsize{5}{10}\selectfont\faStarOfLife}} How mentally demanding was the task? \emph{(1 - Very low, 10 - Very high)}} & \mylabel{\tool ($\mu=4.3 / 10$)} \newline \tlxbarchart{0.2}{0.3}{0.35}{0.1}{0.05}{0}{50\%}{15\%} \\
& \tlxbarchart{0.1}{0.1}{0.45}{0.25}{0.1}{0}{20\%}{35\%} \newline \mylabel{Baseline ($\mu=5.9 / 10$)} \\
\midrule
\multirow{2}{\linewidth}{How physically demanding was the task? \emph{(1 - Very low, 10 - Very high)}} & \mylabel{\tool ($\mu=3.2 / 10$)} \newline \tlxbarchart{0.5}{0.25}{0.20}{0.0}{0.05}{0}{70\%}{05\%} \\
& \tlxbarchart{0.40}{0.05}{0.35}{0.15}{0.05}{0}{45\%}{20\%}  \newline \mylabel{Baseline ($\mu=4.3 / 10$)} \\
\midrule
\multirow{2}{\linewidth}{How hurried or rushed was the pace of the task? \emph{(1 - Very low, 10 - Very high)}} & \mylabel{\tool ($\mu=4.7 / 10$)} \newline \tlxbarchart{0.25}{0.1}{0.50}{0.1}{0.05}{0}{35\%}{15\%} \\
&  \tlxbarchart{0.15}{0.25}{0.35}{0.15}{0.1}{0}{40\%}{25\%}  \newline \mylabel{Baseline ($\mu=5.0 / 10$} \\
\midrule
\multirow{2}{\linewidth}{How successful were you in accomplishing what you were asked to do? \emph{(1 - Perfect, 10 - Failure)}} & \mylabel{\tool ($\mu=4.0 / 10$)} \newline \tlxbarchart{0.2}{0.45}{0.2}{0.15}{0.0}{0}{65\%}{15\%} \\
& \tlxbarchart{0.1}{0.35}{0.30}{0.15}{0.10}{0}{45\%}{25\%}  \newline \mylabel{Baseline ($\mu=5.1 / 10$)} \\
\midrule
\multirow{2}{\linewidth}{How hard did you have to work to accomplish your level of performance? \emph{(1 - Very low, 10 - Very high)}} & \mylabel{\tool ($\mu=5.6 / 10$)} \newline \tlxbarchart{0.05}{0.2}{0.4}{0.35}{0.0}{0}{25\%}{35\%} \\
& \tlxbarchart{0.05}{0.05}{0.4}{0.5}{0.0}{0}{10\%}{50\%}  \newline \mylabel{Baseline ($\mu=6.0 / 10$)} \\
\midrule
\multirow{2}{\linewidth}{\raisebox{1pt}{\textnormal{\fontsize{5}{10}\selectfont\faStarOfLife}} How insecure, discouraged, irritated, stressed, and annoyed were you? \emph{(1 - Very low, 10 - Very high)}} & \mylabel{\tool ($\mu=3.2 / 10$)} \newline \tlxbarchart{0.4}{0.4}{0.15}{0.05}{0.0}{0}{80\%}{05\%} \\
& \tlxbarchart{0.25}{0.2}{0.2}{0.25}{0.1}{0}{40\%}{40\%}  \newline \mylabel{Baseline ($\mu=5.5 / 10$)} \\
\midrule
\multicolumn{2}{c}{\mylegend{1 -- 2\xspace\xspace\xspace}{blue2} \mylegend{3 -- 4\xspace\xspace\xspace}{blue1} \mylegend{5 -- 6\xspace\xspace\xspace}{gray1}\mylegend{7 -- 8\xspace\xspace\xspace}{orange1} \mylegend{9 -- 10\xspace\xspace\xspace}{orange2}} \\
\bottomrule
\end{tabular}
\end{table*}

\subsubsection{Experience of Creating Spreadsheets}
\label{sec:evaluation-user-study-results-experience}
The questionnaire results are summarized in Table~\ref{tab:survey-results}. 
Overall, \tool was rated higher than the baseline across all conversational quality, NASA TLX constructs, and SUS scores. 
The mean SUS score for \tool was 74.0, indicating good usability, compared to the baseline's mean score of 59.8, which suggests OK usability~\cite{bangor2009determining}.
Out of 20 participants, 14 preferred \tool over the baseline, and 16 preferred \tool if response times were equal. 
Below, we discuss participants' experiences while developing spreadsheets with \tool and the baseline, providing insight on the reasons for their preferences.

\paragraph{\textbf{\tool proactively performs actions on its own, while the baseline defers more control to the programmer.}}
Participants observed \tool being more \evalcode{proactive} ($N=9$) in performing actions which \pquote{narrowed down}{11} the problem space and \pquote{took out the need for [programmers] to think}{19}.
This resulted in lower mental ($p<0.05$) and physical ($p=0.12$) demand compared to the baseline (see Table~\ref{tab:survey-results}).
In contrast, the baseline delegated more control to programmers by providing \evalcode{direct manipulation} ($N=11$) to control actions performed on the spreadsheet in the chat window (e.g., accept action and undo buttons).
It also provided \evalcode{more information} ($N=5$) on how to approach the task (e.g., more explanations on how to solve the problem (P9) and multiple example tables (P2)), causing it to generate longer answers and making it difficult to read (P1).
As a result, P11 compared the baseline to a \pquote{Google search engine, just bringing up information that was helpful to a bit}{11}.

The participants expressed different opinions on which approach they preferred. 
Some participants appreciated the approach of \tool in directly executing actions since it was more efficient (P1, P5, P10, P13, P15, P20): \pquote{I find it laborious and a little too repetitious to continually have to tell it, 'Yes, I want you to do what I just told you to do'}{20}.
However, many participants preferred the controls that the \baseline provided in the chat window (P3, P11, P12, P15, P16, P18, P19) since it was important to \pquote{feel like you're in control}{19} to maintain ownership over the spreadsheet.
P20 also suggested the addition of a stop button to terminate the agent's actions midway through.
While P6 and P10 both described the actions of \tool as going \pquote{above and beyond}{6, P10}, only P10 appreciated how it thought about \pquote{those details that are overlooked}{10}.
Meanwhile, P6 felt this was \pquote{invasive}{6} and an \pquote{overreach}{6}: \pquote{If I ask [the AI] something, I feel like I need it to be very specific, and I'd want the tool to also be that specific}{6}.

\paragraph{\textbf{Interactions with \tool were more conversational and collaborative, while interactions with the baseline were more efficient.}}
Participants found \tool to be more \evalcode{collaborative} in its approach ($N=11$), which required conversational interaction throughout developing the spreadsheet.
Meanwhile, study participants described the baseline as more \evalcode{straightforward} in its interaction ($N=3$) and required less programmer input to perform actions in the spreadsheet.
Although several participants felt positively about \tool's approach (P3, P10, P17), others felt that it was cumbersome (P2, P16).
Some participants found \tool to require \pquote{more input}{19} which could make using it more \pquote{taxing}{2} because the programmers \pquote{had to think about what to put}{16} in the spreadsheet.
Other participants found that the questions \tool asked were more \pquote{helpful}{2} and \pquote{engaging}{10} and resulted in personalized spreadsheets (P2, P7).
For the \baseline, participants found its approach more \pquote{technical}{7, P14}, since programmers \pquote{have to know what you're doing to guide [the baseline] where you want it to go}{11}.

\pblockquote{[The baseline] was more of a command interface. It made me feel as though I was giving it things to do, and it was working.
For me, [\tool felt] like we were working collaboratively...
It's going step by step, making sure that I'm on point with every step that it's doing and making sure it's communicating. 
The way the tool was designed did not take anything away from me. 
What it really took away from me was headaches involved with spreadsheet creation and management and data points.}{10}

\paragraph{\textbf{\tool provides better results for more complex tasks, while the baseline performs well for simple tasks.}}
Participants reported feeling more successful in accomplishing the task using \tool compared to the baseline ($p=0.12$, see Table~\ref{tab:survey-results}).
Many participants felt that \tool produced \evalcode{better outcomes} in the main task and performed the required actions better ($N=14$).
Study participants described \tool performing the exact actions they wanted it to do (P6, P10, P18, P19, P20), which made it easier to control or modify the resulting spreadsheet (P11, P12, P14) and perform further actions (P13).
Participants felt that they \pquote{got quite a lot further into the task}{4} and that the spreadsheet \pquote{look[ed] more reliable}{11}.
As a result, P3 noted that this \pquote{eliminates trial and error because it builds it right [the] first [time]}{3}.
Despite the observed \evalcode{lag} ($N=7$), participants felt \tool \pquote{performed beyond...expectations}{8} and was still willing to use the tool:
\pblockquote{Even though [\tool] took a little bit longer for the generation, the quality of the output was so much better that I was quite happy to wait that little bit extra.}{4}

Meanwhile, numerous participants ($N=9$) described instances in which using the baseline resulted in \evalcode{worse outcomes} on the main task.
Study participants noted times where the \baseline did not complete the action the participant wanted (P1, P4, P8, P12, P17).
However, P4, P6, P7, P10, P13 noted that the baseline performed well on \pquote{simple, structured tasks}{4}, such as the tutorial task, and produced a less complex spreadsheet (P7).
As a result, P10 and P13 felt that they needed to break down the problem for the baseline, rather than letting the tool problem solve on its own: \pquote{I found that [with TableTalk] I tried to throw everything all at once...versus [for the baseline] I tried to feed it a little bit less, like one piece at a time of information}{13}.

\paragraph{\textbf{\tool provides better conversational interactions.}}
For conversational quality metrics~\cite{finch-choi-2020-towards} (see Table~\ref{tab:survey-results}), \tool had improved scores in \emph{relevance} ($p=0.009$) and \emph{proactivity} ($p=0.07$) compared to the baseline.
In addition, study participants also noted that \tool provided \evalcode{higher-quality conversation} ($N=9$).
The participants felt that the tool better understood their intentions (P4, P8, P20), making the overall conversation smoother (P7, P14): \pquote{I had a lot of typos in that prompt and it...understood my demands very well}{8}.
Study participants also felt that \tool had \pquote{more humanness}{4} and, as a result, \pquote{on an emotional level...I could understand [it] a lot more}{14}.
In contrast, participants observed that the \baseline provided \evalcode{lower-quality conversation} ($N=10$).
Multiple participants felt that the tool misunderstood their intentions (P3, P8, P9, P11, P18) or had difficulty understanding why the tool performed certain actions without clear explanations (P1, P14), which increased frustration ($p<0.05$) and decreased SUS scores ($p<0.05$) when using the baseline.

\paragraph{\textbf{\tool's suggestions are more relevant to the programmer's context.}}
16 of 20 participants used \tool's suggestions, while 4 participants used the suggestions from the \baseline, implying that \tool suggestions were a useful feature.
Study participants commented on \tool having \evalcode{pertinent suggestions} ($N=14$).
Even if the suggestions were not used in the interaction, the participants felt that they served as a useful reference to drive future interactions (P3, P4, P10, P11, P16) since \pquote{sometimes [\tool's suggestions] gives you things you might not have even thought about}{17}.
One participant also mentioned that the suggestions contained \pquote{language that the AI bot...wanted to use...[to] train me this is the way I word things}{4}.
While \tool's suggestions were generally relevant, P13 felt they could be even more individualized to their personal circumstances.
Meanwhile, study participants commented on the baseline having \evalcode{generic suggestions} ($N=4$), causing them to ignore the suggestions (P1, P3) or find them unhelpful (P9).

\mybox{
\faArrowCircleRight\xspace\textbf{Key findings (RQ7):}
Using \tool affects the programmer's experience by taking a more proactive and collaborative approach, which decreases mental demand compared to the \baseline.
While participants felt that \tool achieves better results in more complex tasks, they had mixed reactions about its approach because it reduces user control in implementing the spreadsheet.
}

\section{Discussion}
\label{sec:discussion}
\tool is a prototype system that demonstrates human-agent collaboration in spreadsheet programming. 
Our results show that language agents can significantly benefit spreadsheet programmers by addressing key challenges they face.
Returning to \tool's design principles (Section~\ref{sec:design-goals}), its flexible scaffolding (DP1, DP2) enabled programmers to focus on the \evalcode{requirements} of the spreadsheet rather than on low-level implementation details by reducing the need to send \evalcode{low-level commands}. 
Additionally, \tool's proactive suggestions alleviated the burden of problem-solving by automatically providing and sending \evalcode{high-level commands}. 
This addressed challenges such as \formativecode{minimal guidance on using advanced Excel features} and \formativecode{limited support on structuring problems}, as identified in the formative study and prior literature~\cite{pirolli2005sensemaking, chalhoub2022s, baltes2018towards, liang2022understanding, ko2011state, nardi1991twinkling}.
Furthermore, the incremental development approach (DP3) resulted in less manual adaptation, addressing templates being \formativecode{hard to adapt}. 
Collectively, these features reduced the mental demand of building spreadsheets and enabled programmers to produce \evalcode{better outcomes}, with higher-quality and more \evalcode{polished} spreadsheets that were preferred over those from the baseline.

Based on our findings from \tool's evaluation study, we discuss the design implications on future human-agent collaboration and spreadsheet programming systems (Section~\ref{sec:design-implications}). 
We then consider the impact of agents on spreadsheet programming, end-user programming, AI-assisted programming, and human-agent collaboration (Section~\ref{sec:agentic-spreadsheet-programming}), and conclude with the limitations of our studies (Section~\ref{sec:limitations}).

\subsection{Design Implications}
\label{sec:design-implications}

\subsubsection{Apply Scaffolding Techniques to Complex Tasks}
While \tool's flexible scaffolding approach improved task outcomes over the baseline's non-scaffolded approach, scaffolding techniques are not a silver bullet for all programming problems.
One challenge was that the \evalcode{proactive} and \evalcode{collaborative} nature of \tool was taxing, especially for simple tasks they could perform themselves:
\pblockquote{Because at the time when [TableTalk] asks questions, it still doesn't know how complex the task is when it starts building. 
For the simple stuff, I would prefer simplistic answers with not many options, kind of like, `Here's what it is. Do you like it? Yes, no.'}{3}
This suggests that scaffolding is most effective for complex problems that spreadsheet programmers struggle to solve without adequate support, such as debugging~\cite{bajpai2024lets}, but may introduce excessive overhead for simpler tasks. 
This aligns with findings on AI programming assistants like GitHub Copilot, which show that programmers often know how to solve the problem and use AI to complete tasks more efficiently~\cite{barke2023grounded, liang2024large}.

Future AI programming systems should balance light-weight interventions, such as AI code completion, with more intensive scaffolding from language agents. 
Future work could develop heuristics and automatic detection methods to determine when each technique is most appropriate. 
Additionally, since novice programmers have specific needs for scaffolding~\cite{latoza2020explicit, arab2022exploratory} that may differ from those of experts', future studies could replicate the evaluation with both expert and novice spreadsheet programmers to better understand their differing support needs.

\subsubsection{Provide Direct Manipulation Interfaces to Control the Agent}
A key challenge identified by participants in \tool was the lack of \evalcode{direct manipulation} interfaces to control the agent, a feature that the baseline offered. 
This was the most frequently cited reason for preferring the baseline, with three participants mentioning it in their rationale.
This indicates that the direct manipulation provided by the Excel canvas alone is insufficient when working with a language agent. 
Programmers desire interfaces to assert control over the agent, such as the ability to stop or undo its actions. 
This aligns with \citet{amershi2019guidelines}'s human-AI interaction guidelines, which highlight the importance of global controls for AI systems.
Therefore, we recommend future human-agent systems to offer direct manipulation interfaces to support stopping and undoing agent actions.

\subsubsection{Make the Level of Proactivity Customizable for the Programmer}
While evaluation study participants appreciated having \evalcode{direct manipulation} interfaces to accept agent actions, several also valued \tool's \evalcode{proactive} approach in performing actions. 
Some felt that direct manipulation interfaces to accept agent actions were repetitive when the action had already been requested.
To address this, human-agent collaboration systems should allow programmers to customize the degree of agent proactivity, enabling a balance between automation and user control based on individual preferences.

Future work should explore automatic methods to determine when direct manipulation interfaces should be offered to accept agent actions.
Participants noted that this preference could vary by task context or programmer expertise.
For example, P15 expressed a strong preference for the direct manipulation interface to accept agent actions as an expert spreadsheet programmer. 
In addition, P6 described business contexts where proactivity would be unwelcome:
\pblockquote{Maybe I'm just speaking for me, for the older users that have got 20 plus years of doing it yourself. They might be like...`No, I'm doing this' [when they receive a suggestion].}{15}
\pblockquote{I wonder, though, if you've got a really defined business and your table is very secure in its formatting. I didn't like the fact that it seemed to make changes to my table because I think I'd probably go crazy if I had a really robust table where I'm just entering for that month, and I don't want changes.}{6}

\subsubsection{Incorporate Features to Evaluate the Agent's Output to Minimize Overreliance}
In the user study, participants engaged in minimal debugging or testing behaviors for both \tool and the baseline. 
However, spreadsheet evaluators identified formula errors in the outputs from both tools, indicating that participants over-relied on AI-generated formulas. 
This reflects a broader concern with AI programming assistants, particularly for novice users~\cite{becker2023programming, kazemitabaar2023novices}.
To address this, future human-agent collaboration systems should incorporate features that help evaluate the AI's output, such as generation quality indicators. 
These features can promote appropriate trust between humans and AI~\cite{wang2024investigating}, encouraging users to critically assess and validate AI-generated content.

\subsubsection{Promote Code Comprehension for Better Debugging}
We observed that a lack of support for code comprehension caused some participants to struggle to debug the generated Excel formulas. 
End-user programmers often use haphazard debugging strategies, which can introduce further errors~\cite{ko2011state}. 
For example, P9 requested \tool to generate a formula that displayed "Excellent" for an overall grade of 100\%. 
Instead of inspecting the formula to identify the error, the participant repeatedly described the desired behavior: \pquote{if score is below 100, do not display excellent}{10}.
In response, \tool generated the formula:
\texttt{=IF(SUM(B2:H2)=100, "Excellent", SUM(B2:H2))},
which displays "Excellent" if the sum of grades equals 100 and the summed scores otherwise. 
However, since the resulting column values did not change, P9 was unaware that \tool had adapted the formula. 
This led P9 to prefer the baseline over \tool, citing that it felt \pquote{more advanced}{9}.
To address this, we recommend that human-agent collaboration systems promote code comprehension by providing detailed explanations of generated formulas whenever they are added to the spreadsheet. 

Future research should explore techniques to enhance code comprehension of spreadsheet formulas. 
Approaches like \citet{ferdowsi2023coldeco}'s COLDECO have tackled this by decomposing formulas into helper columns, highlighting representative rows in a summary table, and providing natural language explanations of functions.
Additionally, as demonstrated by P9’s experience, end-user debugging processes can be ineffective and could benefit from scaffolding tools. 
Future work could investigate whether LLM-augmented scaffolding techniques---similar to \citet{bajpai2024lets}'s ROBIN for traditional programming---might effectively guide users through spreadsheet debugging.

\subsection{Broader Implications}
\label{sec:agentic-spreadsheet-programming}
Below, we discuss the implications of this work on spreadsheet programming (Section~\ref{sec:implications-spreadsheet-programming}), end-user programming (Section~\ref{sec:implications-end-user-programming}), AI-assisted programming (Section~\ref{sec:implications-ai-assisted-programming}), and human-agent collaboration (Section~\ref{sec:implications-human-agent-collaboration}).

\subsubsection{Implications on Spreadsheet Programming}
\label{sec:implications-spreadsheet-programming}
Our work points to the potential of AI assistance in spreadsheet programming, which future work should investigate.
Many evaluation study participants expressed their excitement for proactive AI assistance to improve spreadsheet programming, as they found the automation helpful ($N=6$).
Participants described the time-saving effects (P19) and effort reduction (P17) that AI provided, such as eliminating the need to reference online documentation (P3) and reducing small spreadsheet errors (P10).
While these benefits align with those observed in traditional code completion assistants~\cite{barke2023grounded, liang2024large},
additional research is needed to understand how AI assistance may alter spreadsheet programmer behavior, such as through controlled studies comparing receiving AI assistance to having none.
Beyond the productivity improvements, participants (P8, P15) also noted the educational potential of these tools, which future work could investigate:
\pblockquote{[Having an AI] just asking what you want, would for me personally, free up time for the senior analysts.
You've got a junior analyst...sitting with you, and they're asking all the time, `How do I do this? How do I do that?'
You give them a course to do in Excel...but they're still asking, `How do I do this?'
It's easy now [since] having an [AI] is better at training someone than a training course. [Before AI], if you're training, you have to take it in, write notes, and keep going back.}{15}

\subsubsection{Implications on End-User Programming}
\label{sec:implications-end-user-programming}
Given \tool's ability to assist with spreadsheet programming, our work suggests potential for language agents to help with end-user programming at large, as mentioned in prior work~\cite{reqallyouneed}.
Since \tool users focus on the requirements of the program---which is vital in AI-assisted end-user programming~\cite{reqallyouneed}---rather than on the specific programming notations, end-user programming tools that implement \tool's design principles could reduce many of the documented barriers in end-user programming.
Even with low-code or no-code representations for end-user programming systems~\cite{blackwell2024moral}, end-user programmers have historically struggled with barriers related to using programming notations.
This includes \textit{design} barriers (knowing what should be programmed), \textit{selection} barriers (finding and using the correct programming interfaces), \textit{coordination} barriers (knowing how to compose programming interfaces), and \textit{use} barriers (understanding programming interfaces)~\cite{ko2004six}.
However, while \tool addresses many of the prior barriers in end-user programming, due to the lack of debugging support, the system does not address \textit{understanding} barriers (understanding aspects of the program's external behavior that obscure what actually occurred) and \textit{information} barriers (obtaining information about the program's internal state, like variable values) in end-user programming~\cite{ko2004six}.
Future research is needed to explore how language agents could address these barriers in end-user programming.

\subsubsection{Implications on AI-Assisted Programming}
\label{sec:implications-ai-assisted-programming}
Given that spreadsheet programming is often considered a form of programming~\cite{ko2011state, blackwell2024moral}, our work has implications for AI-assisted programming.
In particular, the participants' experiences with \tool highlight the tensions of working with proactive language agents in programming environments.
While proactive assistance from \tool was seen as helpful by some participants, others expressed concerns that proactive AI assistance might undermine their ownership of the spreadsheet: \pquote{[I want the AI to] make it feel a bit more personal, that it's my co-pilot on my journey, and this is my journey to create this one little project among many others. It's important to me}{19}
This underscores the need for future research into thoughtfully designed proactive coding agent interactions, like how to tailor an agent’s level of proactivity, what information about the user and programming environment to share, when the agent should intervene, and how its presence should be integrated into the programming environment~\cite{pu2025assistance, chen2025need, kumar2025sharp}.

In addition, the findings from using \tool suggests that the design principles could be effective for coding language agents as well.
Work in traditional software development suggests that scaffolding (DP1) is beneficial to programmers~\cite{latoza2020explicit, arab2022exploratory, bajpai2024lets}; the results from \tool indicate that LLMs are an effective way to scaffold programming tasks, as in prior work~\cite{prather2023s, leinonen2023comparing, franklin2025generative, xing2025use}, since it enables flexibility (DP2) within a structured process. 
Recent work also show promise in AI-aided scaffolding in software development, which can increase the quality of AI follow-up suggestions and improve collaboration between programmers and AI programming assistants~\cite{kumar2025sharp, reqallyouneed}.
In addition, incrementally generating code (DP3) could help to obtain higher-quality outputs by regularly obtaining natural language feedback, as \tool users spent less time implementing code themselves and more on describing the requirements of the program.
This incremental approach has also been shown to improve developer collaboration with AI programming assistants~\cite{kumar2025sharp, reqallyouneed}.
Taken together, existing commercial and open-source agents (e.g., Cursor~\cite{cursor2025cursor} or GPT Pilot~\cite{github2025gptpilot}) should consider the design implications above and support flexible scaffolding of complex tasks for programmers.

\subsubsection{Implications on Human-Agent Collaboration}
\label{sec:implications-human-agent-collaboration}
Our work also has implications on human-agent collaboration.
Unlike many prior works that assume full automation and no interactivity between the agent and its user~\cite{liu2024largelanguagemodelbasedagents} (e.g.,~\cite{yang2024swe, li2024sheetcopilot}), our work underscores the importance of human-agent collaboration for open-ended tasks, since the interactive approach of \tool showed promise in improved performance outcomes.
Similar to other work~\cite{feng2024cocoa}, \tool's human-in-the-loop planning showed promise as a human-agent collaboration approach.
A majority of participants used \tool's suggestions directly for human-agent planning.
Even if the suggestions were not accepted, they were still used as inspiration to describe the next step.
By exposing the \tool user to multiple potential next steps, the user and agent collaboratively explored different problem-solving steps to improve task outcomes, similar to how having an LLM explore different problem-solving approaches improves performance~\cite{yao2024tree}.
The benefits of human-agent collaboration corroborates existing work, which finds that agents that interact with humans achieve better performance compared to non-interactive agents by relying on human expertise to disambiguate queries~\cite{vijayvargiya2025interactive}.

Given the nascency of agents, further research is needed to better understand human-agent collaboration, such as understanding the underlying design challenges and guidelines for human-agent interactions~\cite{bansal2024challenges}.
While \tool investigates how humans and agents can collaborate, a key question largely overlooked by this and other work is what effective human-agent interaction interfaces should look like. 
Future studies should explore new interaction paradigms for human–agent collaboration beyond natural language chat interfaces

\subsection{Limitations}
\label{sec:limitations}
We now discuss the limitations of \tool, the template study, and formative and evaluation user studies.

\paragraph{Impact of \tool on Spreadsheet Programming}
The current design of \tool serves as a scaffolding tool to assist spreadsheet programmers in creating spreadsheets. 
We do not claim that all spreadsheet programming tasks should be fully automated. 
Instead, we advocate for a collaborative approach where LLMs assist programmers in tackling the most challenging aspects of spreadsheet programming, a benefit demonstrated by the evaluation study results.
However, as with any LLM-based technology, potential negative impacts deserve further investigation. 
AI programming assistants like GitHub Copilot can generate code that is difficult to understand or debug~\cite{liang2024large}.
While the evaluation of \tool shows promise in offsetting the added metacognitive load associated with using generative AI~\cite{tankelevitch2024metacognitive}, at its extreme, it could cause spreadsheet programmers to use less critical thinking, which has been observed with generative AI usage~\cite{lee2025impact}.
Thus, spreadsheet programmers may face similar challenges, such as overlooking critical errors generated by \tool or experiencing reduced learning due to overreliance on the tool.
However, we anticipate the effect of decreased critical thinking could be reduced in \tool, as the tool encourages users to engage more with their domain expertise by focusing more on spreadsheet requirements.
\citet{lee2025impact} suggests that having users engage with domain knowledge could be effective in mitigating the reduction in critical thinking in knowledge workers.
Future work should explore how to design tools that manage these risks to ensure that these tools effectively support users without undermining their skills or decreasing critical thinking.

As more agent-based approaches are adopted for spreadsheet programming, it is also important to consider how to scale \tool to real-world spreadsheets since in practice, spreadsheets can be large.
Currently, \tool may not scale to these scenarios, since the tool is given the full spreadsheet as context in the prompt, and further, regenerates the spreadsheet data in full with the {\texttt{create\_table}} tool (see Table~\ref{tab:tools}).
To scale the current approach, future research could explore providing 
compressed spreadsheet representations~\cite{dong2025spreadsheetllmencodingspreadsheetslarge} to \tool
rather than the full spreadsheet,
or defining tools that allow the agent to insert, delete, or modify individual rows instead of regenerating the full spreadsheet.

\paragraph{Study Limitations}
Our studies have several limitations.
For internal validity, the research team likely had confirmation biases that could influence the qualitative analyses. 
We mitigated this threat by involving multiple coders to develop the codebook and measuring inter-rater reliability.
In addition, the results of the formative study could be influenced by the participants' lack of training with spreadsheet templates.
We anticipate this threat was reduced in the study because the participants had experience with spreadsheet templates, with four of them reporting spreadsheet template usage in their work.
For the evaluation study, participants interacted with \tool through the researcher's computer rather than on their personal machine; this change in the participant's spreadsheet development environment could affect the study results.
Further, the \tool interface supported both the spreadsheet canvas and chat, leaving less space for the spreadsheet canvas on lower-resolution screens.
Thus, the results of the study may not generalize to all computing environments, such as higher-resolution screens.

In terms of external validity, for the template study, the template selection process and the smaller sample may have introduced biases. 
Thus we do not claim that our sample of templates is representative of all spreadsheets. 
Instead, this study serves as the foundation for the design of \tool.
We reduced the threat of developing the codebook on a single template domain by performing an additional validation on 23 spreadsheet templates from a range of domains and found no new codes.
Meanwhile, for all user studies, participants completed tasks under time pressure, which may not be representative of all working conditions.
Also, user study participants---English speakers from the United States, Canada, Australia, and the United Kingdom---may not be fully representative of all spreadsheet programmers, especially due to the smaller sample sizes of each user study.
For the \tool evaluation, we focused on two tasks, but these may not encompass the full range of spreadsheet programming activities. 
While our findings provide insights into spreadsheets and AI-assisted programming, these factors limit the generalizability of our results.
Future work could explore the effects of \tool in different contexts, with more complex tasks or larger-scale datasets, such as through a longitudinal field study. 

\paragraph{Researcher Positionality}
As researchers, our lived experiences and positionality influence our perspectives and, consequently, how we approach our research~\cite{savin2023qualitative, holmes2020researcher}. 
Below, we outline aspects of our positionality that are most relevant to this study.
The research team comprises computer science researchers, including full-time scientists, research fellows, and research interns on the Excel team, with expertise in human-computer interaction, software engineering, and artificial intelligence. 
We approach spreadsheet tools from software-oriented, user-centered, and machine learning perspectives, which may lead to an underemphasis on cultural or sociological aspects of spreadsheet use.
While our close ties to Excel and its research and development provide a deep understanding of spreadsheet programming tools, we acknowledge that this proximity can influence our experiments and introduce implicit biases in how we frame Excel's strengths and limitations. 
Additionally, our affiliation with Excel might create a power dynamic in user studies, making participants reluctant to provide critical feedback.
To address potential courtesy bias, we requested participants to discuss the potential limitations of each tool.

\section{Conclusion}
Spreadsheet programming is a challenging activity that requires programmers to schematize data, apply problem-solving skills to break down complex tasks, and effectively use APIs to implement spreadsheet formulas.
Research has shown that LLMs can guide programmers through complex tasks through scaffolding approaches, which re-structure and decompose tasks in a way such that users can accomplish tasks independently.
However, to our knowledge, no prior work has explored scaffolding specifically for spreadsheet programming.

In this work, we address the challenges of spreadsheet programmers by leveraging the increasing capabilities of LLM-based language agents to solve complex problems and generate spreadsheet formulas.
We introduce \tool, a language agent-based system that helps spreadsheet programmers create spreadsheets, informed by a user study of 7 spreadsheet programmers and a study of 85 Excel spreadsheet templates.
\tool implements a flexible scaffolding approach, using language agents to help programmers build spreadsheets incrementally.
It has access to tools (i.e., external modules that are used to complete specific operations) to help implement atomic spreadsheet features that can be composed to build spreadsheets.
We find that compared to the \baseline, \tool enables programmers to build more polished spreadsheets and reduces cognitive load.
It allows programmers to focus more on the requirements of the spreadsheet rather than on the problem solving and implementation details.
Therefore, \tool serves as a proof-of-concept that LLM-based scaffolding is a promising avenue to support spreadsheet programmers through complex tasks.
To facilitate replication of this work, we include the protocols for the formative and evaluation user studies, the spreadsheet template dataset, all codebooks used across the analyses, and a video demo of \tool in the supplemental materials~\cite{supplemental-materials}.

\begin{acks}
We thank our study participants for their insights.
We also thank Albert Dodson for technical assistance in implementing \tool, as well as Brad A. Myers and Mukhul Singh for valuable insight into the project.
In addition, we extend our gratitude to Soham Pardeshi, Millicent Li, Christopher Kang, and Michelle Lin for the feedback on \tool.
Last but not least, we give special thanks to Mei \meiicon, an outstanding canine software engineering researcher, for providing support and motivation throughout this study.
Jenny T. Liang conducted this work as an Applied Scientist Intern in Microsoft’s PROSE Team (\url{https://www.microsoft.com/en-us/research/group/prose}).
She is supported by the National Science Foundation under grants DGE1745016 and DGE2140739.
\end{acks}

\bibliographystyle{ACM-Reference-Format}
\bibliography{acmart}

\appendix

\section*{Appendix Overview}
To provide context on the study protocols used in the formative (Section~\ref{sec:formative-user-study}) and evaluation (Section~\ref{sec:evaluation-user-study}) studies, we have included a small subset of the protocols in the appendix in Sections~\ref{app:formative-user-study} and~\ref{app:evaluation-user-study} respectively for the reader's convenience.
To access all of the study materials, including the spreadsheet template dataset, full study protocols, codebooks, and a video demo of \tool, please refer to the materials that are publicly available on FigShare~\cite{supplemental-materials}.

\section{Formative User Study}
\label{app:formative-user-study}

\subsection{Food Drive Task Tracking Instructions}
You are working with your colleagues, Arya and Zarah, to coordinate project tasks for your local food drive. Your goal is to understand who is doing the most work between you three for the food drive. Here are the tasks that need to be tracked and who is assigned to them.
\begin{enumerate}
    \item Collecting food donations: Done once a week by Zarah and Arya. This takes 3 hours to complete.
    \item Coordinating volunteers: Done by you every 2 weeks. This takes an hour to complete.
    \item Counting food inventory: Done by you every two days. This task takes 45 minutes to complete.
    \item Donation box placement: Done by you once quarterly. This takes about 3 hours to complete.
    \item Donation delivery: Done twice a week by Arya. This task takes about 30 minutes to complete. 
    \item Local food bank distribution planning: Done once a week by Zarah. This takes about an hour.
    \item Local shelter distribution planning: Done once a week by Arya. This takes about an hour.
    \item Marketing on social media: Done by Arya once a month. This takes 2 hours to complete.
\end{enumerate}

\subsection{Design Probe}
Imagine you had a system that could help you create Excel tables from scratch. Suppose this system had access to your working context, such as your files and your browser. Also suppose that this system could generate Excel formulas correctly every time.

\subsubsection*{Prototype Screenshots}
\begin{figure}[t!]
\begin{center}
\includegraphics[trim=0 175 0 175, clip, width=\linewidth, keepaspectratio]{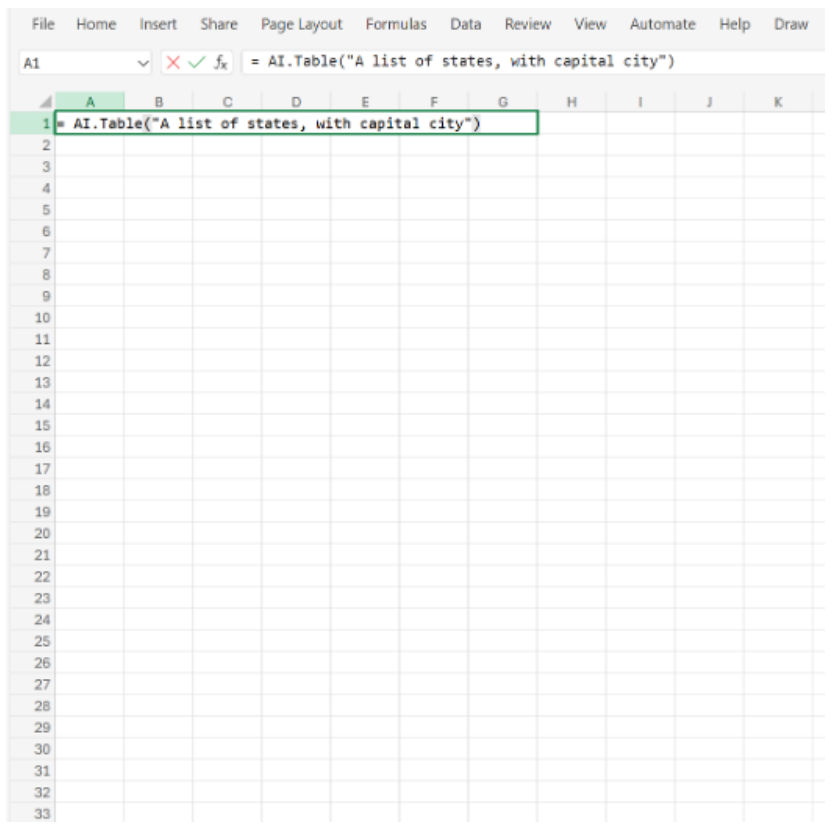} 
\end{center}
\caption{A screenshot shown to the formative participants during the design probe for Option 1: Code snippets on the Excel canvas.}
\label{fig:prototype1}
\end{figure}

\begin{figure}[t!]
\begin{center}
\includegraphics[trim=0 150 1025 0, clip, width=0.75\linewidth, keepaspectratio]{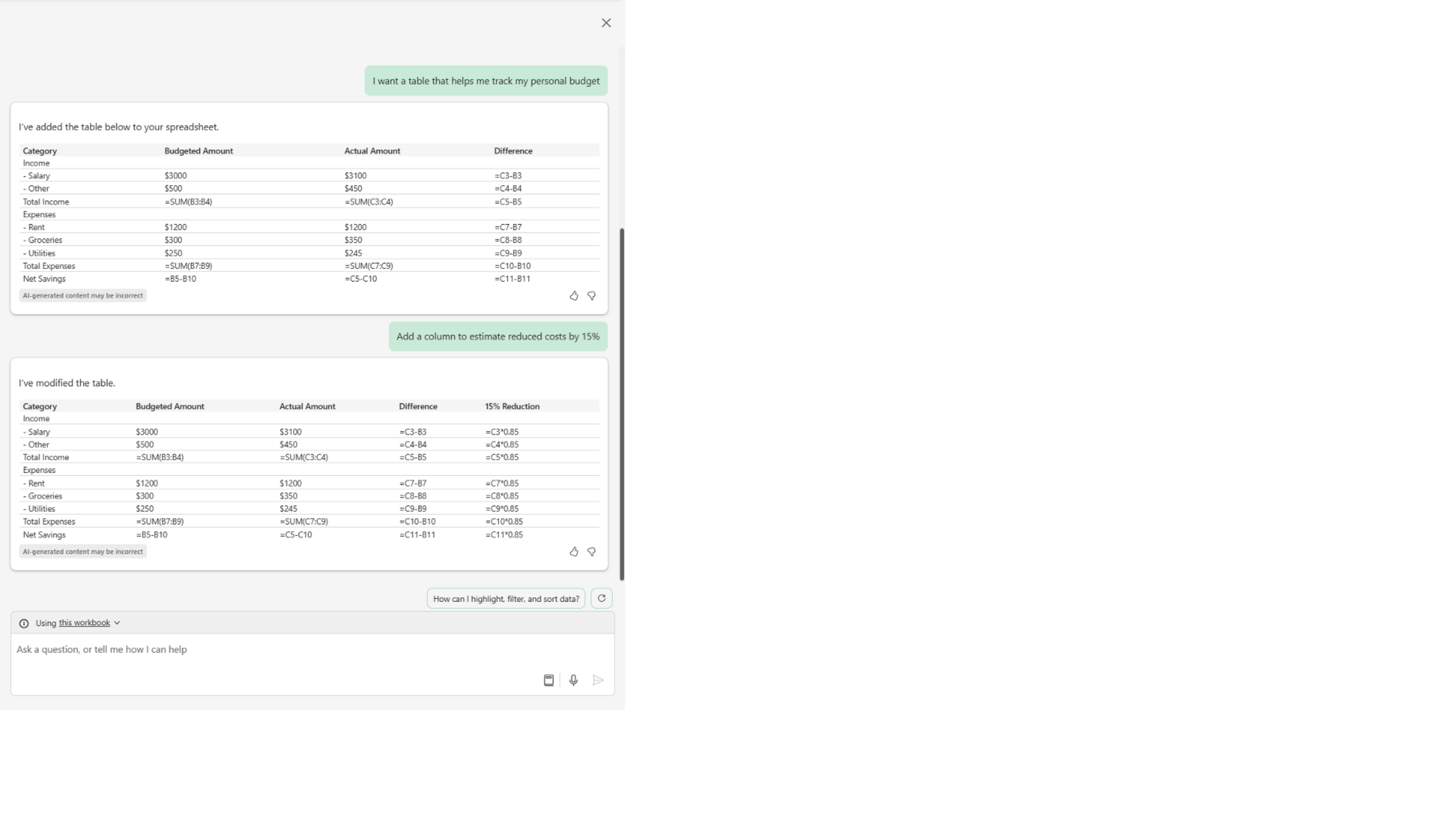} 
\end{center}
\caption{A screenshot shown to the formative participants during the design probe for Option 2: Chat with an AI chatbot.}
\label{fig:prototype2}
\end{figure}

\begin{figure}[t!]
\begin{center}
\includegraphics[trim=0 150 1175 0, clip, width=0.75\linewidth, keepaspectratio]{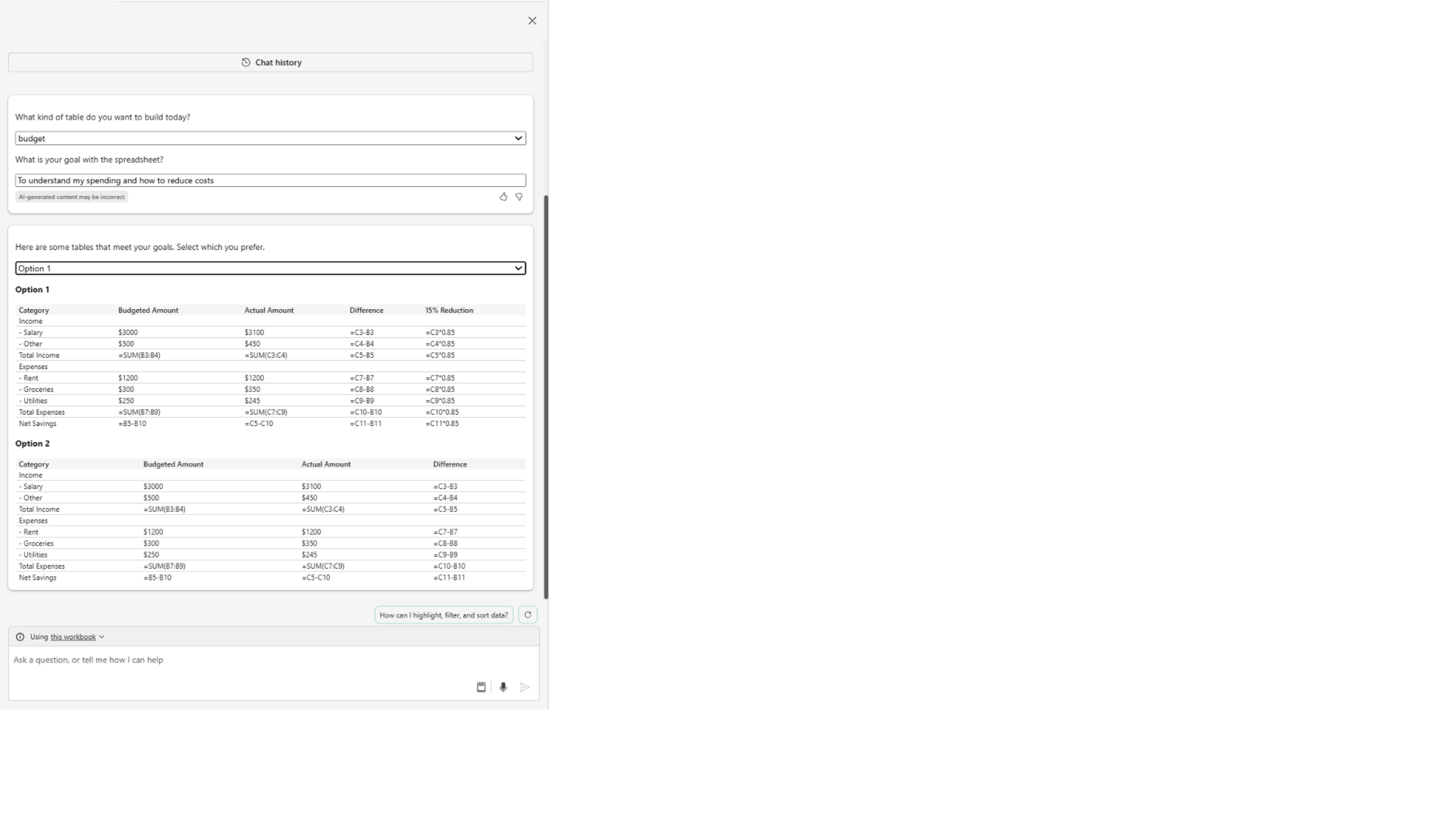} 
\end{center}
\caption{A screenshot shown to the formative participants during the design probe for Option 3: Setup wizard to scaffold table creation.}
\label{fig:prototype3}
\end{figure}

To view the screenshots of the different prototypes that were  to participants during the design investigation in the formative study, refer to Figures~\ref{fig:prototype1},~\ref{fig:prototype2}, and~\ref{fig:prototype3}.

\subsection{Template}
\begin{figure}[h!]
    \centering
    \begin{subfigure}{\textwidth}
        \includegraphics[trim=0 275 0 25,clip,width=\textwidth]{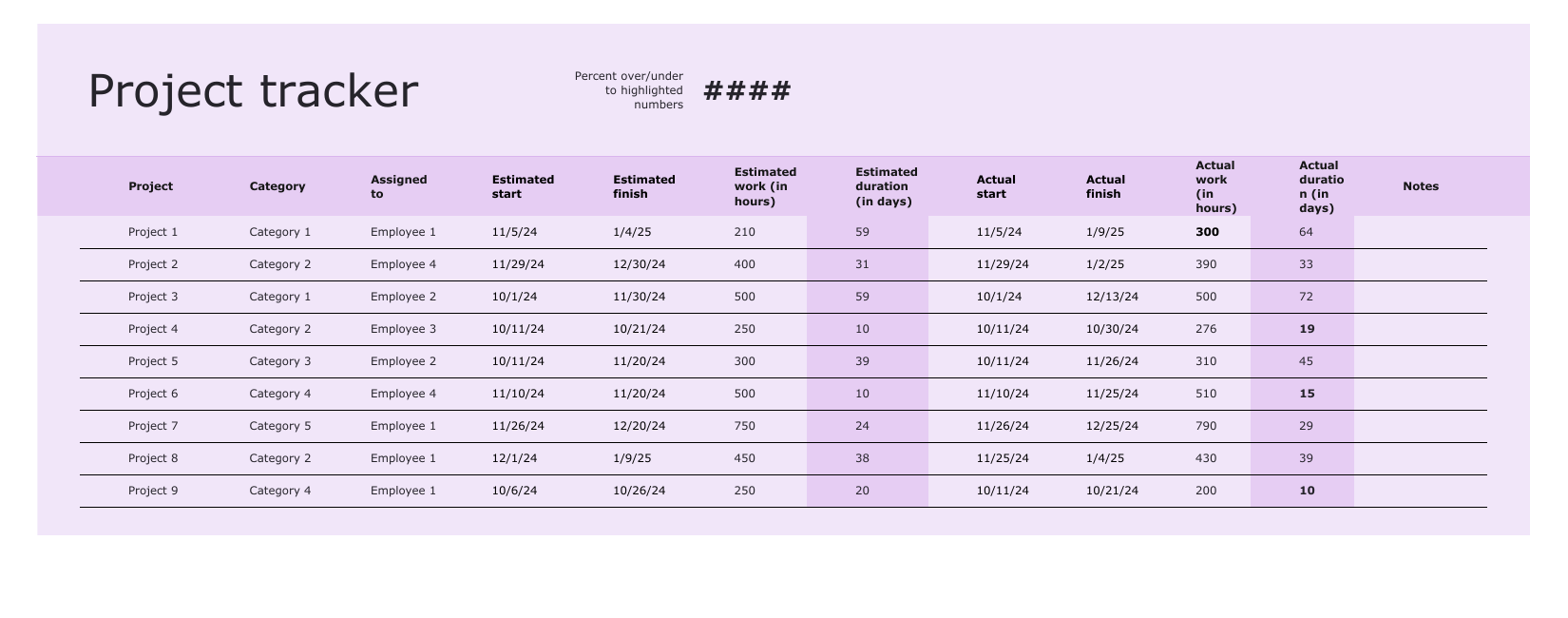}
    \end{subfigure}
    \begin{subfigure}{0.5\textwidth}
        \includegraphics[trim=25 250 150 75,clip,width=\textwidth]{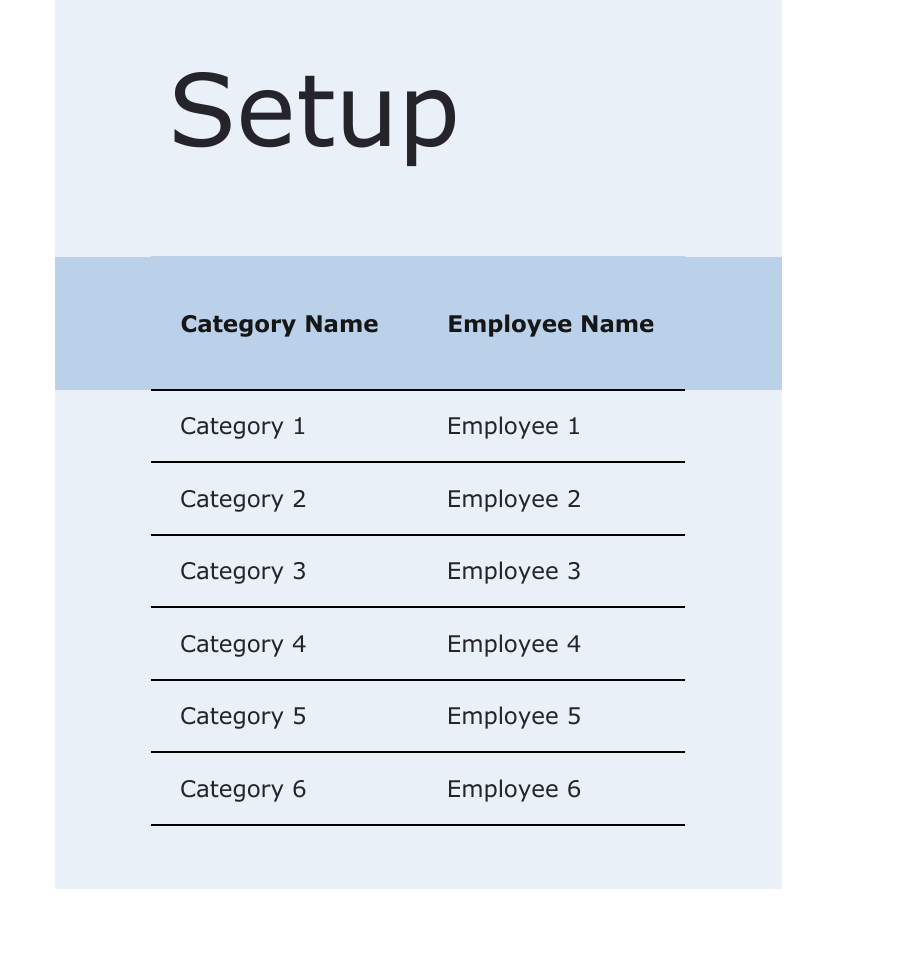}
    \end{subfigure}

    \caption{
        A screenshot of the Excel spreadsheet template used during the formative user study.
        The template contained two sheets: one titled "Project Tracker" (top) and one titled "Setup" (bottom).
    }
    \label{fig:spreadsheet-template}
\end{figure}

To view the screenshots of the template that was used during the formative user study, refer to Figure~\ref{fig:spreadsheet-template}.

\section{Evaluation User Study}
\label{app:evaluation-user-study}

\subsection{Student Grades Analysis Instructions}
\subsubsection*{\textbf{Instructions}} Work with the AI tool to develop a spreadsheet that addresses the given scenario. You can think of yourself as a spreadsheet programmer, and the moderator as your client for the spreadsheet. 
While developing the spreadsheet, you should ask questions about the requirements of the spreadsheets. You can think of your own questions or use the results from the AI to think of questions to ask.

\subsubsection*{\textbf{Scenario}}
You are a teacher who taught a course in geometry to 3 high school students. Your job is to help the principal (the moderator) count how many times each of the overall grades occur (e.g., the number of students who earned an "Excellent" overall grade) based on graded items in the course.

\subsubsection*{\textbf{Details}} The scenario includes the following details.
\begin{itemize}
    \item In your course, there were 4 assignments, 2 exams, and 1 participation activity. 
    \item Grades are expressed as a percentage.
    \item The assignments are worth 40\% of the grade; the exams are worth 50\% of the grade; the participation points are worth 10\% of the grade. 
    \item The overall grades go from Excellent (95+\%), Very good (85\%+), Satisfactory (75\%+), Needs improvement (below 75\%).
\end{itemize}

\subsubsection*{\textbf{Data}} 
There is existing data for the grades of the 3 students. You can ask the principal (moderator) for the specific grades once you work with the AI to develop a table and are satisfied with its schema.

\subsubsection*{\textbf{Spreadsheet quality}} Your final spreadsheet should be: 1) polished enough for a coworker to reuse with minimal effort to get the same functionality on different data and 2) be visually appealing.

\subsection{Hours Worked \& Client Payments Tracking Instructions}

\subsubsection*{\textbf{Instructions}}
Work with the AI tool to develop a spreadsheet to address the given scenario. You can think of yourself as a spreadsheet programmer and the moderator as your client for the spreadsheet. 
While developing the spreadsheet, you should ask questions about the requirements of the spreadsheets. You can think of your own questions or use the results from the AI to think of questions to ask.

\subsubsection*{\textbf{Scenario}}
You are an assistant to an entrepreneur (the moderator). The entrepreneur needs help calculating how much money she earned from different kinds of tasks done in September (e.g., earning \$500 from consulting-related work) by working with clients.

\subsubsection*{\textbf{Details}} 
The scenario includes the following details.
\begin{itemize}
    \item The entrepreneur’s work has different hourly rates based on the service provided. 
    \item When the job takes more than 10 hours, each hour past the first 10 hours costs 10\% more than the hourly base rate. 
    \item Some clients also receive discounts at a flat amount (e.g., a \$100 discount).
    \item The services provided were completed on different dates.
\end{itemize}

\subsubsection*{\textbf{Data}}
There is existing data for the 4 clients. You can ask the entrepreneur (moderator) for the specific data once you work with the AI to develop a table and are satisfied with its schema.

\subsubsection*{\textbf{Spreadsheet quality}}
Your final spreadsheet should be: 1) polished enough for a coworker to reuse with minimal effort to get the same functionality on different data and 2) be visually appealing.

\end{document}